\documentclass[letterpaper]{article}
\usepackage{array}
\usepackage{tabularx}
\usepackage[T1]{fontenc}

\usepackage{geometry}
\usepackage{setspace}

\usepackage[style = chem-acs]{biblatex}
\usepackage{graphicx}
\usepackage{float}
\newfloat{scheme}{htbp}{los}
\floatname{scheme}{Scheme}
\floatname{chart}{Chart}
\newfloat{graph}{htbp}{loh}

\usepackage{chemformula} 
\usepackage[version = 4]{mhchem} 

\usepackage{authblk}
\author[1]{Ruiqing Zang}
\author[1]{Jianan Xiao}
\author[1]{Jacob Bair}
\author[1]{Ian Wang}
\author[1]{Chuyue Peng}
\author[1]{Matthias Kuehne*}
\affil[1]{Department of Physics, Brown University, Providence, RI 02912, USA}

\title{Filling and Sealing Carbon Nanotubes by Local Anodic Oxidation}
\date{*Email: kuehne@brown.edu}

\begin{document}

\maketitle

\begin{abstract}
Carbon nanotubes (CNTs) are prime nanofluidic conduits and nanopores for studying matter under extreme confinement. Since most synthesis routes yield CNTs with closed ends, opening and filling strategies are required to access their interior for fluid and ion transport, whereas studies of the encapsulated species themselves may additionally require sealing to prevent their escape in dry or vacuum environments. Capacitively coupled local anodic oxidation (LAO) in an atomic force microscope (AFM) has been shown to cut individual CNTs at predefined positions, but whether cutting induces opening, filling, or sealing has remained unclear. Using spatially resolved micro-Raman spectroscopy, we find that LAO cutting surprisingly produces filled CNT segments that remain sealed under dry Ar atmosphere for weeks. The magnitudes of the blueshift and broadening of the innermost-wall radial breathing-like mode are consistent with water filling, pointing to the water meniscus condensed between the AFM tip and the substrate as the source of CNT filling during cutting. High-vacuum annealing at 140~°C releases the encapsulated species from some, but not all, segments, indicating a thermally surmountable barrier rather than re-formed end caps. This one-step cut-fill-seal capability offers a route to diameter-resolved studies of nanoconfined water and aqueous solutions, and to zeptoliter-scale sample delivery for vacuum-based analytical techniques.
\end{abstract}

There is significant interest in carbon nanotubes (CNTs) for their ability to take up various atomic and molecular substances \cite{Pederson1992,Ajayan1993,Dujardin1994,Campo2021}, which positions them as prime model systems for the study of matter under extreme confinement \cite{Aluru2023}, as containers or templates for chemical reactions \cite{Ugarte1996,Allard2024}, and as pores for fluidic and ionic transport \cite{Aluru2023, Li2025}. As most synthesis routes yield CNTs with closed ends, opening procedures are of great importance to enable access to the hollow CNT interior. The various methods for opening CNTs can be divided into physical and chemical approaches. Physical routes include breaking nanotubes through ultrasonication in solution \cite{Wenseleers2007,Campo2021} and mechanical milling such as with a focused ion beam \cite{Kuehne2021,Tu2024}. Reported chemical routes include oxidation in air \cite{Ajayan1993b}, heating in carbon dioxide \cite{Tsang1993}, immersion in acids such as nitric acid \cite{Tsang1994,Faucher2021}, and oxygen plasma treatment \cite{Lee2010,Liu2010,Faucher2021}. Many of these approaches are global, exposing the entire sample to the same conditions. For chemical routes then, selective opening relies solely on the higher reactivity of the CNT end caps compared to the sidewalls, but the sidewalls may nonetheless be attacked, particularly at defects. For substrate-bound CNTs, such unwanted damage can be confined through lithographic masking \cite{Lee2010,Liu2010,Faucher2021}, but only at the cost of additional processing steps and resist contact with the CNTs. Direct-write techniques that open individual CNTs at predefined positions without masks are therefore highly desirable.

Beyond opening and filling, studies of matter confined in the CNT interior may additionally require sealing, so that the encapsulated species remain inside. This is particularly critical for vacuum-based analytical techniques, where volatile encapsulated species are expected to escape through open CNT ends, as noted for water in electron microscopy \cite{Naguib2004,Xu2026}. Water encapsulation has been reported after hydrothermal treatment of closed multi-walled CNTs \cite{Naguib2004} and after laser-induced oxidation of suspended single-walled CNTs \cite{Chiashi2014}, but filling yields were low or unquantified, and the mechanisms by which water enters and remains inside were proposed rather than established. More recently, vibrational electron energy-loss spectroscopy resolved hydrogen bonding of water inside individual CNTs, but required focused-ion-beam cutting, exposure to near-saturated humidity, and manual epoxy sealing of the cut regions, with only about 13~\% of examined tubes showing a water signal \cite{Xu2026}. A method that opens, fills, and seals individual CNTs at predefined positions in a single step would greatly simplify such studies.

Scanning probe-based techniques offer a maskless route to cutting individual CNTs at predefined positions \cite{Venema1997,Hertel1998,Park2002,Hyon2005,Kumar2011,Li2018}. Among these, local anodic oxidation (LAO) is particularly versatile. Initially demonstrated on metal and semiconductor surfaces using a scanning tunneling microscope \cite{Thundat1990,Dagata1990}, LAO is now typically implemented in an atomic force microscope (AFM) and has received increasing attention as a nanolithographic etching technique, particularly for graphene \cite{Giesbers2008,Weng2008,Masubuchi2009,Puddy2011,Byun2011,Li2018,Wu2025}. LAO relies on water splitting within a water meniscus that condenses between the AFM tip and the substrate (Fig.~1a), driven by extreme local electric fields exceeding $10^9$~V/m (1~V/nm) \cite{Sugimura1993,Morimoto1997,Weeks2005,Butt2011}. Although AFM-based LAO has been shown to cut CNTs locally \cite{Kumar2011,Li2018}, reports remain scarce, and whether LAO cutting also opens, fills, or seals the CNT has not been addressed.

Here, we study the opening, filling, and sealing of individual CNTs by capacitively coupled AFM-based LAO. We first confirm that LAO reliably cuts CNTs under a distinct set of operating parameters, reinforcing the generality of the technique beyond the limited existing reports \cite{Kumar2011,Li2018}. To probe opening, we monitor the water-filling-induced blueshift and broadening of the radial breathing(-like) mode (RBM) of the innermost wall of individual CNTs \cite{Wenseleers2007,Cambre2010,Faucher2021,Tu2024}. Water filling is a particularly stringent and sensitive probe of opening, as capillary filling of CNTs with water occurs spontaneously down to at least a 0.55~nm diameter \cite{Wenseleers2007,Cambre2010,Faucher2021,Campo2021,Tu2024}, so that a lack of a filling signal after cutting reflects a cut end that is not genuinely open rather than a diameter-dependent filling limitation. We find that cutting does not necessarily open a CNT. Surprisingly, however, all four CNTs examined by Raman spectroscopy in this work, selected for a highly resonant innermost-wall RBM at 532~nm excitation, yielded filled and sealed segments upon LAO cutting, as evidenced by a persistent RBM blueshift and broadening in dry atmosphere that is quantitatively consistent with water filling. While the physical nature of the seal remains unclear, high-vacuum annealing at 140~$^\circ$C empties some segments, a subset of which then remains open, whereas others remain filled even after annealing, up to at least 16 days after cutting. We discuss the potential of this approach for zeptoliter-scale sample delivery and the ingredients of capacitively coupled LAO that may underlie it, pointing to the tip–substrate meniscus chemistry as a key subject for future study.

\begin{figure}
  \centering
  \includegraphics[width=\textwidth]{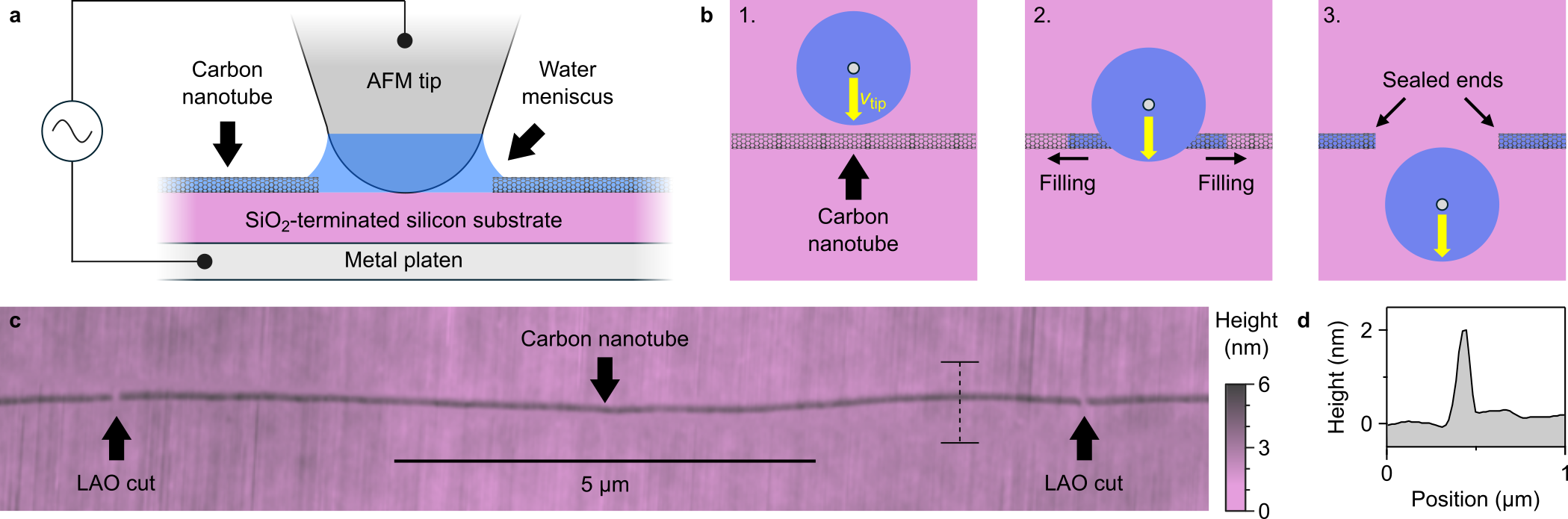}
  \caption{Capacitively-coupled LAO of CNTs. (a) Schematic of the experimental setup and water meniscus. (b) Top view schematic of the LAO cutting procedure. (c) AFM image after LAO cutting of CNT~1. (d) Height profile along the black dashed line in (c).}
  \label{fig:1}
\end{figure}

Figure~1 summarizes the AFM-based LAO procedure we use to locally cut individual CNTs. Prior to CNT growth, nonstochiometric SiO$_x$ markers are deposited on SiO$_2$-terminated Si substrates by optical lithography, SiO$_2$ sputtering, and lift-off, allowing us to locate the same CNT in Raman spectroscopy and AFM. CNTs are then grown by a low-gas-flow variant of an established ``kite-growth'' chemical vapor deposition method \cite{Huang2003,Jin2007}. Briefly, a drop of 0.1~M FeCl$_3$ in ethanol is deposited at the upstream end of the substrate, which is heated to 900~$^\circ$C in a tube furnace under N$_2$/H$_2$ (100/8~sccm). Growth proceeds at 970~$^\circ$C for 45~min under CH$_4$/H$_2$ (4/8~sccm), followed by cooling to room temperature under N$_2$/H$_2$ (300/8~sccm). This yields sparse arrays of typically single-, double-, and triple-walled CNTs that may be partially bundled and reach millimeter lengths \cite{Tu2024}. We perform LAO in a Bruker Dimension Icon AFM housed in an enclosure with a domestic air humidifier and a relative humidity ($RH$) meter. A Pt/Ir-coated (95/5~\%) tip is brought into contact with the substrate, where a nanoscale water meniscus forms by capillary condensation (Figure~1a); its spatial extent defines the electrochemical reaction region. To cut, we translate the tip laterally across the target CNT in constant-force mode (feedback on) while applying an AC bias between the tip and the metal platen beneath the sample, which couples capacitively across the SiO$_2$-terminated Si substrate (Figure~1b; Table~\ref{tbl:1}; Supporting Information section S1). Unless otherwise noted, we confirm each LAO cut by tapping-mode AFM imaging immediately afterwards, which reveals a gap separating two otherwise continuous CNT sections. If no gap is visible, we repeat the LAO procedure. Figure~1c shows a representative CNT (``CNT~1'’) after two cuts spaced approximately 10~µm apart (black arrows), which divide it into three segments. The height profile in Figure~1d indicates an outer diameter of approximately 2~nm, representative of all CNTs studied here.

\begin{table*}[t]
  \caption{CNT LAO cutting parameters.}
  \label{tbl:1}
  \centering
  \footnotesize
  \setlength{\tabcolsep}{3pt}
  \renewcommand{\arraystretch}{1.2}

  \begin{tabularx}{\textwidth}{
    |>{\raggedright\arraybackslash}p{0.14\textwidth}
    |*{5}{>{\raggedright\arraybackslash}X|}
  }
    \hline
    \textbf{Parameter}
    & \textbf{Value for CNT~1}
    & \textbf{Value for CNT~2}
    & \textbf{Value for CNT~3}
    & \textbf{Value for CNT~4}
    & \textbf{Value for CNT~5} \\
    \hline

    Tip
    & \multicolumn{5}{c|}{Pt/Ir-coated non-contact AFM tip (Arrow NCPt)} \\
    \hline

    Force
    & \multicolumn{5}{c|}{3~$\mu$N} \\
    \hline

    Scan speed
    & 200~nm/s
    & 100~nm/s
    & 100~nm/s
    & 100~nm/s
    & 200~nm/s \\
    \hline

    Scan mode
    & \multicolumn{5}{c|}{constant force (feedback loop on)} \\
    \hline

    Voltage setting
    & $16~\mathrm{V_{pp}}$ sinusoidal ac voltage, 100~kHz, no dc offset
    & $10~\mathrm{V_{pp}}$ sinusoidal ac voltage, 100~kHz, $-3$~V dc offset
    & $16~\mathrm{V_{pp}}$ sinusoidal ac voltage, 100~kHz, no dc offset
    & $16~\mathrm{V_{pp}}$ sinusoidal ac voltage, 100~kHz, no dc offset
    & $10~\mathrm{V_{pp}}$ sinusoidal ac voltage, 100~kHz, $-3$~V dc offset \\
    \hline

    Relative Humidity
    & $>50~\%$
    & $>50~\%$
    & $50~\%$
    & $70~\%$
    & $50~\%$ \\
    \hline

    Average success rate per cut
    & $100~\%$
    & $75~\%$
    & $50~\%$
    & $100~\%$
    & $75~\%$ \\
    \hline
  \end{tabularx}
\end{table*}

\begin{figure}
  \centering
  \includegraphics[width=6.5in]{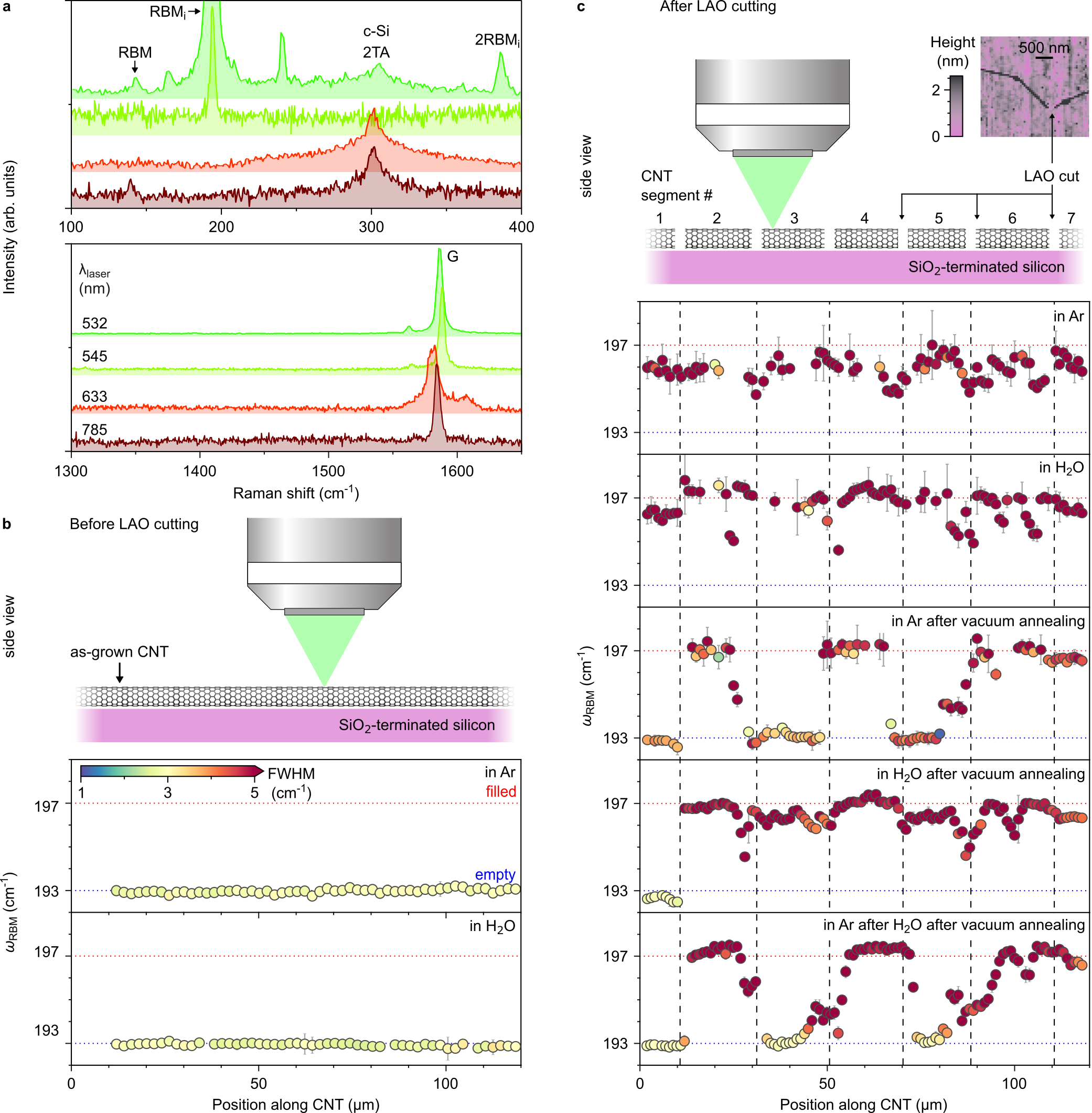}
  \caption{(a) Micro-Raman spectra of CNT~2 measured at different Raman excitation wavelengths $\lambda_\mathrm{laser}$ as indicated. (b-c) Micro-Raman spectroscopy measured along CNT~2 (b) before and (c) after LAO cutting and in different environments as indicated (in chronological order from top to bottom). Top panels are schematic side views of the experiment. Bottom panels show frequency, $\omega_\mathrm{RBM}$, and full width at half maximum, FWHM, of the most resonant RBM observed at $\lambda_\mathrm{laser}=532~\mathrm{nm}$ that we assign to the innermost wall of a few-walled tube (see panel a). Black dashed lines in (c) indicate approximate LAO cut locations. Blue and red dotted lines in (b-c) indicate $\omega_\mathrm{RBM}$ of the empty and filled state, respectively.}
  \label{fig:2}
\end{figure}

To assess whether LAO cutting also opens CNTs and enables interior filling, we exploit the RBM of the innermost wall of few-walled CNTs, which is highly shielded from the external environment by the outer wall(s) \cite{Pfeiffer2003,Rao2007,Faucher2023} and thus serves as a sensitive probe of interior filling \cite{Wenseleers2007,Faucher2021}. We select CNTs exhibiting a narrow RBM peak, with a linewidth of only a few cm$^{-1}$, approaching the $\sim1~\mathrm{cm}^{-1}$ width reported for inner walls \cite{Pfeiffer2003,Rao2007}. The selected RBM must also be strongly resonant at $\lambda_\mathrm{laser}=532~\mathrm{nm}$, and there must be no additional RBM peaks at higher frequencies from the same tube. We denote this mode RBM$_\mathrm{i}$. Two additional observations support its assignment to a shielded innermost wall. First, before cutting, immersing the CNT in water induces no measurable shift or broadening of RBM$_\mathrm{i}$ compared to Ar, showing that this wall is insensitive to exterior water and therefore unlikely to be the outermost wall. Second, RBM$_\mathrm{i}$ blueshifts and broadens by several cm$^{-1}$ only after LAO cutting and only in some segments, as expected for a wall in direct contact with interior water.

Figure~2a shows representative micro-Raman spectra of one such CNT (``CNT~2''), acquired at laser excitation wavelengths $\lambda_\mathrm{laser}$ as indicated (see Supporting Information Section~S2 for more details on the Raman spectroscopy). Besides the 2TA phonon feature of silicon (c-Si) from the substrate, the spectra show RBM$_\mathrm{i}$ at $193~\mathrm{cm^{-1}}$, which is highly resonant at $\lambda_\mathrm{laser} = 532$ nm and narrow as well (FWHM $\approx3~\mathrm{cm}^{-1}$). The empirical relation for pristine single-walled CNTs, $d\approx228~\mathrm{nm\cdot cm^{-1}}/\omega_\mathrm{RBM}$ \cite{Liu2011}, suggests an approximate diameter of $\sim1.2~\mathrm{nm}$ (owing to inter-wall coupling \cite{Levshov2017}, the actual diameter of this wall is likely somewhat larger, but its exact value is not important here since it is, in any case, above the threshold for spontaneous filling). Spectra acquired at 532~nm and 785~nm laser excitation wavelengths further reveal a broader mode at $140~\mathrm{cm}^{-1}$ stemming from a larger-diameter wall. Given its small frequency separation from RBM$_\mathrm{i}$, assigning it to a concentric outer wall would imply a substantial upshift from coupling to the exterior environment (substrate, adsorbates) \cite{Pfeiffer2003,Rao2007,Levshov2011,Levshov2017,Tu2024}. Alternatively, it may stem from a different CNT in a bundle. The G band near $1580~\mathrm{cm}^{-1}$ comprises at least four components, which also indicates the presence of at least 2 walls. At $\lambda_\mathrm{laser} = 532$ nm, we additionally observe the RBM$_\mathrm{i}$ overtone at $386~\mathrm{cm}^{-1}$ and weak peaks at $166$ and $240~\mathrm{cm}^{-1}$. As their resonance tracks that of RBM$_\mathrm{i}$, we tentatively assign them to combination bands, possibly involving inter-wall modes \cite{Luo2008,Lim2014,Tailor2018}. If the $240~\mathrm{cm}^{-1}$ peak were instead an RBM, its corresponding diameter ($\sim0.95~\mathrm{nm}$) could not fit inside the wall giving rise to RBM$_\mathrm{i}$, so it would have to stem from another CNT in a bundle. The spectra are therefore consistent with our assignment of RBM$_\mathrm{i}$ to the innermost wall. Additionally, the absence of a defect-induced D band near $1350~\mathrm{cm}^{-1}$ attests to the high crystalline quality of CNT~2, typical of our ultralong CNTs.

To establish a baseline for our measurements, we first record Raman line scans on the as-grown CNT~2 sequentially in Ar and deionized (DI) water at $\lambda_\mathrm{laser}=532~\mathrm{nm}$ (Figure~2b). Line scans are obtained by translating the sample with a motorized stage so that the laser spot steps along the CNT axis. The frequency $\omega_\mathrm{RBM}$ and FWHM of RBM$_\mathrm{i}$ are extracted by Lorentzian fitting after referencing each spectrum to the Si peak at $520.7~\mathrm{cm}^{-1}$. Only fits meeting specific quantitative quality criteria are retained (Supporting Information Section~S2). As-grown CNTs are typically closed and do not take up water upon immersion \cite{Faucher2021}. Consistent with this expectation, $\omega_\mathrm{RBM}$ remains at $193~\mathrm{cm}^{-1}$ with a FWHM of $\sim3~\mathrm{cm}^{-1}$ along the entire measured length in both environments. This spatially uniform response provides a well-defined reference for changes induced by LAO cutting.

We then LAO-cut CNT~2 at six locations with a pitch of $\sim20$~µm, using the parameters stated in Table~\ref{tbl:1}. AFM images (inset to Figure~2c; Supporting Information Section S3) reveal successful LAO cuts and in some cases slight tube displacement within the first 1-2~µm around the cut location. Raman line scans recorded sequentially in Ar and DI water (Figure~2c) reveal that, in both environments, RBM$_\mathrm{i}$ has blueshifted by $\sim4~\mathrm{cm}^{-1}$ to $\omega_\mathrm{RBM}\approx197~\mathrm{cm}^{-1}$ and broadened by a factor of $\sim2$ in all segments of CNT~2, in good agreement with reports on water filling of $\sim1.2~\mathrm{nm}$-diameter single-walled CNTs which show 2-4~cm$^{-1}$ blueshifts depending on chirality and approximately doubling of the linewidth \cite{Cambre2010,Fagan2011,Faucher2021}. The possibility that LAO-induced sidewall damage led to these changes is unlikely: the D band remains unchanged throughout the experiment, and quantitative Raman mapping after similar LAO cutting of graphene shows no change in defect density only 2~µm away from the cut (Supporting Information Section~S4). Surprisingly, however, the filling signature persists in dry Ar, in which water would be expected to escape through open CNT ends. Its retention thus requires some form of sealing at the cut sites.

To further investigate the sealing of the CNT segments and the retention of encapsulated water, we anneal the sample at $140\,^{\circ}\mathrm{C}$ for $8\,\mathrm{h}$ under high vacuum ($3\times10^{-6}\,\mathrm{mbar}$). Subsequent Raman line scans in Ar show that $\omega_\mathrm{RBM}$ of segments 1, 3, and 5 has returned to $\sim193~\mathrm{cm^{-1}}$ with a FWHM of $\sim3~\mathrm{cm}^{-1}$, matching the as-grown CNT (Figure~2b,c), indicating that these segments have emptied. This reversibility further confirms that the post-cutting shift and broadening arise from removable interior content rather than permanent structural changes. In contrast, RBM$_\mathrm{i}$ of segments 2, 4, 6, and 7 remains upshifted and broadened, characteristic of the filled state. The G band, which was also recorded at each position and for every condition tested, shows no corresponding segment-dependent changes in the frequency or FWHM of its dominant component (Supporting Information Section~S5), leaving RBM$_\mathrm{i}$ as a clear signature of interior filling. LAO cutting therefore not only divides the CNT into physically separate segments but can also fill and seal them.

We then test whether the segments that were emptied by annealing remain open by immersing the sample in water again (Figure~2c). RBM$_\mathrm{i}$ of segments 3 and 5 blueshifts and broadens as expected for water filling \cite{Cambre2010,Fagan2011,Faucher2021}, reaching nearly the same $\omega_\mathrm{RBM}$ and FWHM as immediately after LAO cutting. This further indicates that water is likely the dominant species entering the CNT during LAO. In contrast, RBM$_\mathrm{i}$ of segment 1 remains unchanged, indicating a closed end, which we tentatively attribute to clogging, as frequently encountered in nanopores. A final line scan in Ar shows that segments 3 and 5 empty again, recovering the $\omega_\mathrm{RBM}$ and FWHM of the as-grown CNT. This directly confirms that water escapes from open CNTs in dry Ar, and hence that the segments retaining water after cutting must be sealed. This final scan, acquired 16 days after LAO cutting, also demonstrates that LAO-filled segments can remain sealed over practically relevant timescales.

\begin{figure}
  \centering
  \includegraphics[width=6.5in]{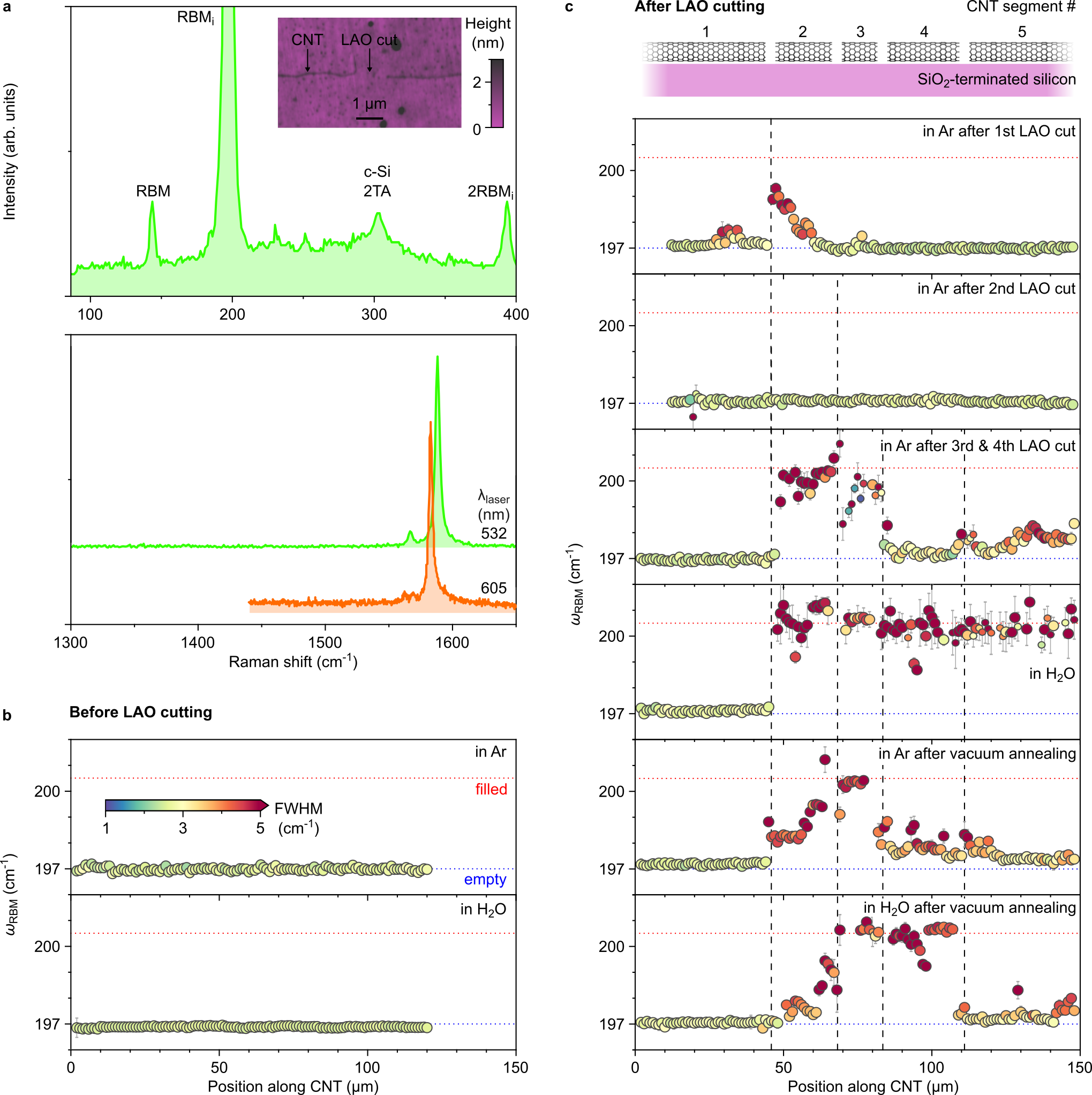}
  \caption{(a) Micro-Raman spectra of CNT~3 measured at different Raman excitation wavelengths $\lambda_\mathrm{laser}$ as indicated. The inset shows the corresponding AFM height profile. (b-c) Micro-Raman spectroscopy measured along CNT~3 (b) before and (c) after LAO cutting and in different environments as indicated (in chronological order from top to bottom). Top panels are schematic side views of the experiment. Bottom panels show frequency, $\omega_\mathrm{RBM}$, and full width at half maximum, FWHM, of the most resonant RBM observed at $\lambda_\mathrm{laser}=532~\mathrm{nm}$ that we assign to the innermost wall of a few-walled tube (see panel a). Black dashed lines in (c) indicate approximate LAO cut locations. Blue and red dotted lines in (b-c) indicate $\omega_\mathrm{RBM}$ of the empty and filled state, respectively.}
  \label{fig:3}
\end{figure}

\begin{figure}[p]
  \centering
  \includegraphics[
    width=\linewidth,
    height=0.8\textheight,
    keepaspectratio
  ]{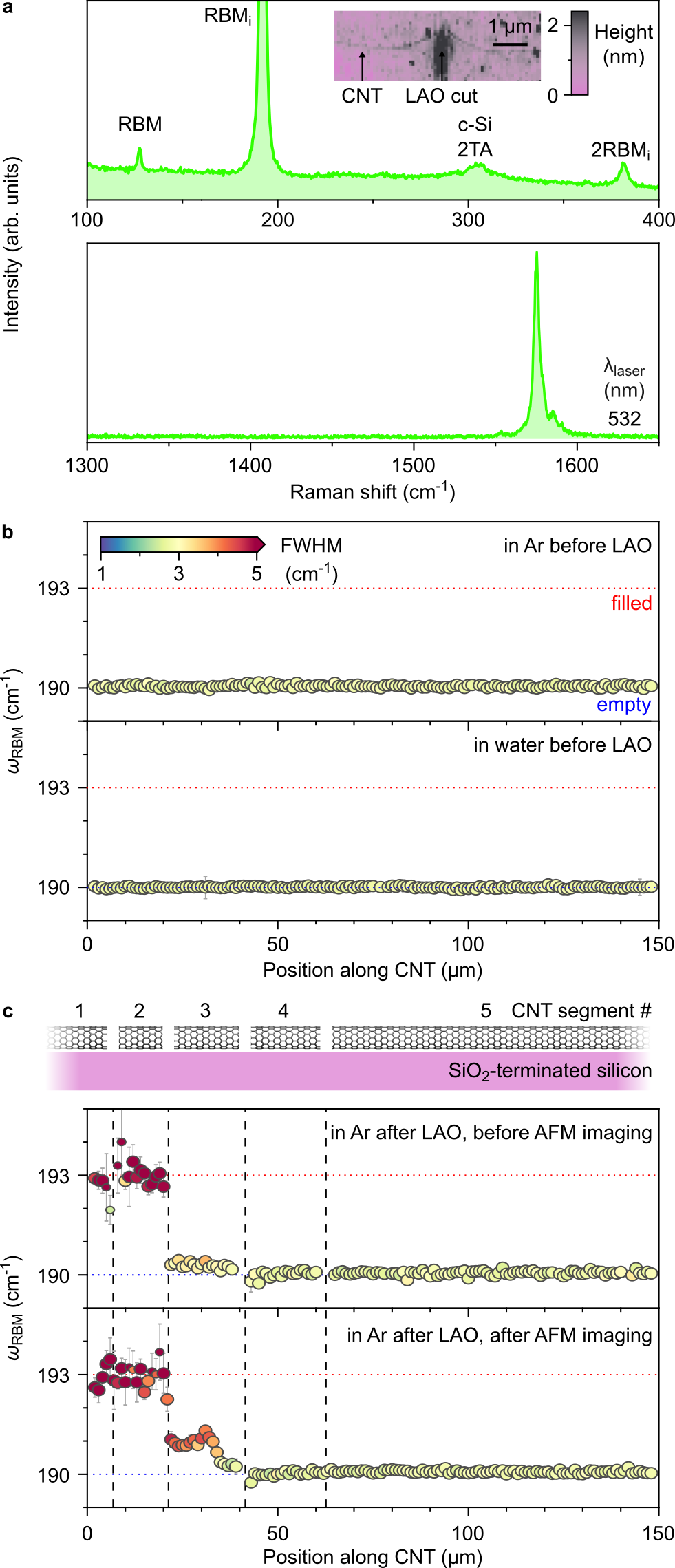}
  \caption{(a) Micro-Raman spectra of CNT~4 measured at $\lambda_\mathrm{laser} = 532$ nm. The inset shows the corresponding AFM image indicating the cut position. (b-c) Micro-Raman spectroscopy measured along CNT~4 (b) before and (c) after LAO cutting, as well as (b) in different environments and (c) before/after AFM tapping-mode imaging as indicated (in chronological order from top to bottom). The frequency, $\omega_\mathrm{RBM}$, and FWHM of the highest frequency RBM stemming from the innermost wall of this CNT are shown. Black dashed lines in (c) indicate approximate LAO cut locations that are also indicated in the schematic side view of the segmented sample in the top panel. Blue and red dotted lines in (b-c) indicate $\omega_\mathrm{RBM}$ of the empty and filled state, respectively.}
  \label{fig:4}
\end{figure}

To examine the repeatability of the LAO-induced response, we performed successive LAO cuts on an additional tube ("CNT~3") and mapped the frequency of the assigned RBM$_\mathrm{i}$ (Figure~3). The first cut produced a localized RBM upshift in segment~2, which disappeared after the second cut, leaving the frequency close to the initial value of approximately $197~\mathrm{cm}^{-1}$. Following the third and fourth cuts, pronounced upshifts were observed in segments~2 and~3, approaching the frequency assigned to the filled state. The reappearance of this spectroscopic signature is consistent with repeated LAO-induced changes in the filling state of the same CNT. Unexpectedly, segment~2 exhibits the filling signature only after cuts 3 and 4, although cut 2 had already physically separated it from the part of the CNT on which these cuts were performed. The reappearance of the filling-associated RBM signature in segment~2 may be partly attributable to tapping-mode imaging. As we typically image each cut by tapping-mode AFM under the same humidity conditions used for LAO, where adsorbed water layers are present on the surface, this raises the question of whether sealing may also occur during imaging rather than during cutting alone.

In an attempt to answer this question, we LAO-cut another CNT (``CNT~4'') at four locations without subsequent tapping-mode imaging, and record Raman line scans in Ar before and after a single tapping-mode scan (Figure~4). CNT~4 shows RBM$_\mathrm{i}$ at $190~\mathrm{cm}^{-1}$, again highly resonant at 532~nm, and a lower-frequency mode at $130~\mathrm{cm}^{-1}$ (Figure~4a). Already before imaging, segments 1 and 2 exhibit the filling signature, demonstrating that LAO cutting alone suffices to fill and seal CNT segments. Subsequent tapping-mode imaging preserves these features but induces a slight upshift in segment~3, consistent at most with partial filling \cite{Faucher2023}. Tapping-mode imaging in a humid environment may therefore also seal water into CNT segments on occasion, which warrants further investigation. G-band line scans and AFM images for CNTs 3 and 4 are provided in Supporting Information Sections~S5 and~S3, respectively, and RBM data for an additional CNT (``CNT~5'') is provided in Section~S6.

Taken together, these results establish capacitively coupled LAO as a direct-write route that cuts, fills, and seals individual CNTs in a single step, at predefined positions and without masks, resists, or electrical contacts to the tube. Its working principle is simple: the tip, the condensed meniscus, and the electrically floating CNT form a transient electrochemical cell driven by displacement current across the substrate, with the CNT alternating between anode and cathode in every ac cycle (Supporting Information Section~S1). This picture turns the LAO parameters into experimental handles. The dc offset and waveform set the oxidation and reset half-cycles and thereby the chemistry at the tip and within the meniscus (Supporting Information Section~S8). The capacitance of the CNT to the substrate, set by its length, should limit the charge transferred per cycle. Measuring the current during cutting would potentially allow monitoring reaction progress directly. Segment length and cutting speed may further tune the filling fraction toward partially filled states hosting nanoconfined liquid–vapor interfaces \cite{Faucher2023,Advincula2026}, while deliberate control of the meniscus composition could extend the approach to filling CNTs with ions, which has so far remained challenging. Notably, a subset of segments remained filled even after annealing at 140~$^\circ$C for 8~h in high vacuum, so the seals formed by LAO can withstand the sample preparation and measurement conditions of vacuum-based analytical techniques. LAO thus provides a route to zeptoliter-scale sample delivery for such techniques, without the humidity exposure and manual sealing that currently limit them \cite{Xu2026}. Establishing the physical nature of the seal and the electrochemistry of the meniscus (Supporting Information Section~S7) will be central to further developing this one-step cut-fill-seal process into a general platform for diameter-resolved studies of water and aqueous solutions under extreme confinement.

\section*{Data Availability}
The data underlying this study are openly available in Zenodo at https://doi.org/10.5281/zenodo.24120335 \cite{Zang2026}.

\section*{Author Contributions}
\textbf{Conceptualization}: R.Z., J.X., J. B., M. K.; \textbf{Data curation}: R.Z., J.X., J. B., I. W.; \textbf{Formal analysis}: R.Z., J.X., J.B., I.W.; \textbf{Funding acquisition}: M.K.; \textbf{Investigation}: R.Z., J.X., J.B., I.W., C.P.; \textbf{Methodology}: R.Z., J.X., J.B., I.W., C.P., M.K.; \textbf{Project administration}: M.K.; \textbf{Resources}: M.K.; \textbf{Software}: R.Z., J.X., J.B., M.K.; \textbf{Supervision}: J.X., M.K.; \textbf{Validation}: R.Z., J.X., J.B., I.W., M.K.; \textbf{Visualization}: R.Z., J.X., J.B., I.W., M.K.; \textbf{Writing - original draft}: R.Z., J.B., I.W., M.K.; \textbf{Writing - review \& editing}: R.Z., J.X., J.B., I.W., C.P., M.K.

\section*{Supporting information}

The Supporting Information is available free of charge.
\begin{itemize}
  \item Capacitively coupled LAO of CNTs. Micro-Raman spectroscopy and spectral analysis. AFM data of CNTs 2-4. Quantitative assessment of LAO-induced defect creation. Micro-Raman spectroscopy data of the G band of CNTs 2-4. Micro-Raman spectroscopy data of CNT 5. Meniscus geometry and electric field distribution. EDX characterization of the AFM tip before and after LAO.
\end{itemize}

\section*{Acknowledgements}

The authors thank Joy Zeng for insightful comments on the electrochemistry. This material is based upon work supported by the National Science Foundation under Grant No. 2341781. I. W. acknowledges support from the Brown University Summer/Semester Projects for Research, Internship, and Teaching (SPRINT) program through an Undergraduate Teaching and Research Awards (UTRA) fellowship.

\printbibliography




  
  

\end{document}



\begin{center}

{\large \textbf{Supporting Information of}}

\vspace{0.5cm}

{\LARGE \textbf{Filling and Sealing Carbon Nanotubes by Local Anodic Oxidation}}

\vspace{0.8cm}

Ruiqing Zang$^{1}$,
Jianan Xiao$^{1}$,
Jacob Bair$^{1}$,
Ian Wang$^{1}$,
Chuyue Peng$^{1}$,
and Matthias Kuehne$^{1,*}$

\vspace{0.6cm}

$^{1}$Department of Physics, Brown University, Providence, Rhode Island 02912, USA

\vspace{0.4cm}

$^{*}$Corresponding author: kuehne@brown.edu

\end{center}

\vspace{1cm}




\tableofcontents
\newpage




\section{Capacitively coupled LAO of CNTs}

\begin{figure}
  \centering
  \includegraphics[width=3.33in]{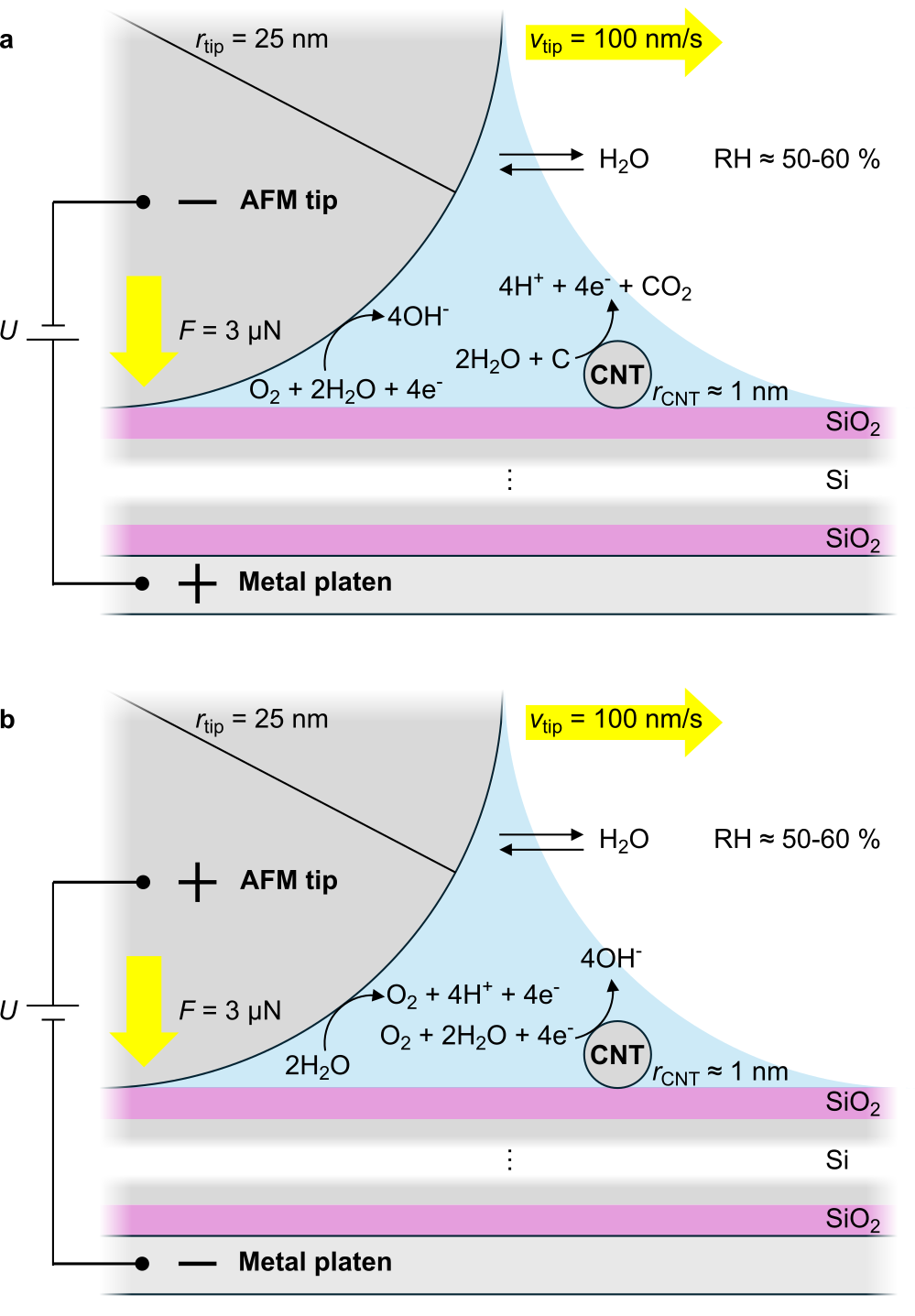}
  \caption{Schematic of the LAO mechanism. (a) First half-cycle with the CNT as the anode and the tip as the cathode. (b) Second half-cycle with the CNT as the cathode and the tip as the anode.}
  \label{fig:5}
\end{figure}

To aid in understanding the conditions of CNT cutting, filling, and sealing, we show schematics of the proposed LAO mechanism in Figure~\ref{fig:5}. The AFM tip, water meniscus, and CNT essentially form an electrochemical cell. Although the CNT is not directly contacted with an electrical lead, it is capacitively coupled to the metal platen via the SiO$_2$-terminated silicon substrate on which it rests, allowing electrochemical reactions to occur at both electrodes transiently upon polarization via displacement current across the silicon substrate. Sustaining the relevant electrochemical reactions beyond a single transient period therefore requires AC operation as we do in this work. During a single AC period, there are two half-cycles: (i) where the CNT acts as the anode and the tip acts as the cathode (Figure~\ref{fig:5}a), and (ii) vice versa (Figure~\ref{fig:5}b). The most relevant reaction occuring at the CNT is the oxidation of carbon during the first half-cycle which proceeds as
\[
\mathrm{C + 2H_2O \rightarrow CO_2 + 4H^+ + 4e^-},
\]
i.e., the local removal of carbon from the CNT and its conversion to carbon dioxide (and possible intermediates). In addition to carbon removal, this anodic reaction generates excess electrons that remain on the CNT because it is electrically floating, i.e., not directly connected to an external electronic reservoir. The electrical charge accumulating on the CNT raises its electrochemical potential. The reaction will therefore stop once the potential of the CNT approaches that of the tip. During the second half-cycle where the CNT serves as the cathode, the excess electrons accumulated on the CNT can then be consumed by oxygen reduction. Oxygen reduction is also the most thermodynamically likely reaction occuring at the AFM tip during the first half-cycle (see Figure~\ref{fig:5}a), as well as possibly metal oxidation. During the second half-cycle, oxygen evolution most likely occurs at the AFM tip as indicated in Figure~\ref{fig:5}b.

Electrochemical reactions at the CNT and the AFM tip cease if the two electrodes come into direct physical contact, which equilibrates their electrochemical potentials through direct electron transfer. This not only can prevent cutting on the first attempt but can also displace the CNT on the substrate \cite{Hertel1998}, which we occasionally observe (see insets of main text Figs.~3a and 4a, as well as Section~\ref{S:AFM} below). Cutting is therefore favored if the water meniscus is significantly taller than the CNT diameter, as sketched in Figures~\ref{fig:5}a–b, such that many AC cycles are completed while the meniscus connects the CNT and the tip before the tip apex reaches the CNT. To assess whether this condition is plausibly met, we estimated the meniscus geometry from the Kelvin equation and computed the resulting electric field distribution by finite-element calculations (Supporting Information Section~S7). At $RH\approx50$–$60~\%$, the zero-field meniscus height remains at best comparable to typical CNT diameters, even when accounting for pre-wetted surfaces. It would also imply cut widths comparable to the tip radius, well below the $\sim0.1-1~\mathrm{\mu m}$ we observe. Electric-field-enhanced capillary condensation can plausibly stabilize a meniscus large enough to account for the observed cut widths. However, the condensed meniscus also defocuses the electric field into a ring around the tip apex \cite{Kremmer2003,Cambel2007}, which requires additional ingredients to reach the $\sim1~\mathrm{V/nm}$ needed to drive water splitting and carbon oxidation. Anisotropic dielectric properties of confined and interfacial water, as well as electric double layers at the solid–liquid interfaces, therefore appear key to a more complete description of the process \cite{Avouris1997,Bonthuis2011,Fumagalli2018,Tran2023,Wang2025}.

Such a description would also place the choice of LAO parameters on a more mechanism-informed footing. As shown in other LAO work \cite{Graf2006}, the modulation frequency, the ratio of reset to oxidation durations, and the oxidation and reset potentials may be exploited as largely independent handles on LAO outcome. With a sinusoidal bias as used here, the dc offset already shifts both the fraction of each period during which the CNT acts as the anode and the potential reached during the reset half-cycle. It could thus be used to suppress tip oxidation (Supporting Information Section~S8) and, more broadly, to control the chemistry at the tip and within the meniscus. Rectangular waveforms, not available on our instrument, would allow the durations and potentials of the oxidation and reset half-cycles to be set independently, rather than jointly through a single offset. Monitoring the current during LAO would further provide a direct measure of reaction progress. Because the CNT is electrically floating, its capacitance to the platen, set by CNT length and dielectric thickness, should limit the charge transferred per half-cycle, so oxidation should subside sooner for shorter segments. Cutting with purely dc bias may therefore succeed only for sufficiently long or few-walled CNTs, which would offer an empirical, local means of distinguishing CNT lengths in the AFM. Segment-to-segment variations in the RBM blueshift moreover suggest differences in filling fraction. Tuning segment length and cutting speed may therefore favor partially filled states, which have been observed in lithographically segmented CNTs \cite{Faucher2023} and are of interest for engineering nanoconfined liquid–vapor interfaces. Systematic statistics across these parameters, including the applied force, should also reveal which conditions favor sealing. Elucidating the nature of the seal will be important in this regard, while the thermally induced emptying of some segments suggests a route to triggered release, for example by in situ heating in an electron microscope. Electrochemical kinetics models, for instance based on the equivalent circuit in Figure~S1, would provide important insight into what dictates the meniscus chemistry and the LAO conditions, including the concentration and nature of charged species such as OH$^-$, H$^+$, carbonate species, and possibly solvated metal ions from tip oxidation. Deliberately tuning the meniscus chemistry could in turn enable filling of CNTs with ions, which has remained challenging.

\subsection{Equivalent electrical circuit for capacitively coupled LAO}

\begin{figure}[H]
    \centering
    \includegraphics[width=3.565in]{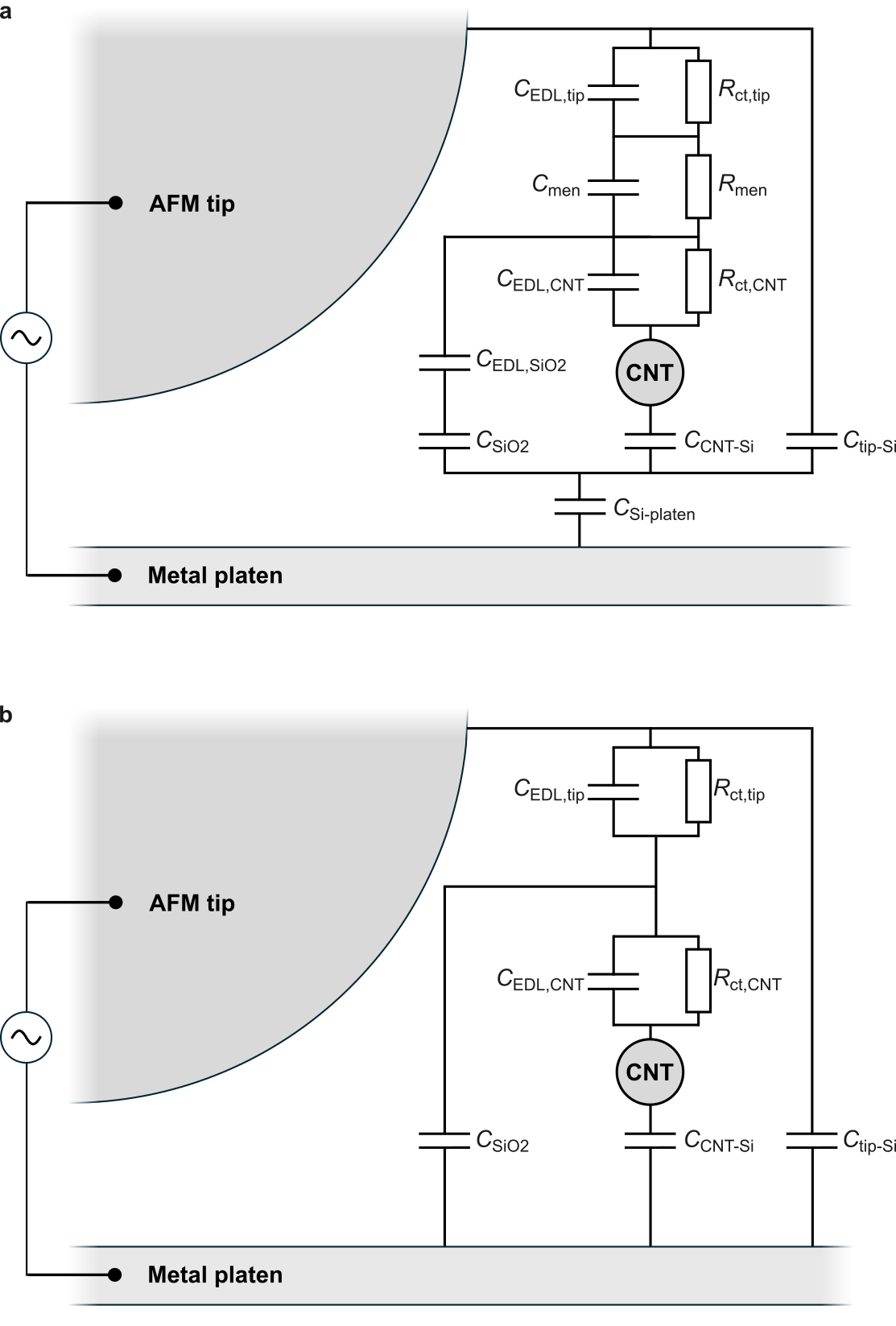}

    \caption{Equivalent electrical circuit for capacitively coupled LAO. (a) More detailed and (b) simplified version as detailed in the text.}
    \label{fig:S1}
\end{figure}

Fig.~\ref{fig:S1}a shows an equivalent electrical circuit for the capacitively coupled LAO geometry discussed in the main text. Dimensional estimates suggest that several elements may be omitted, leading to the simplified circuit shown in Fig.~\ref{fig:S1}b, as follows. The AC voltage is applied between the AFM tip and the metal platen. The degenerately doped bulk of the Si substrate acts as a floating metal electrode that is coupled to the platen through the back oxide, $C_\mathrm{Si-platen}=\varepsilon_0\varepsilon_\mathrm{ox}A_\mathrm{sub}/h_\mathrm{SiO_2}\approx 6$~nF, where $\varepsilon_0$ is the vacuum permittivity, $\varepsilon_\mathrm{ox}=3.9$, $h_\mathrm{SiO_2}\approx 300$~nm is the thickness of both the top and back oxides, and $A_\mathrm{sub}\approx 10~\mathrm{mm}\times 5~\mathrm{mm}$. A meniscus of radius $r_\mathrm{men}$ connects the tip (radius $r_\mathrm{tip}=25$~nm) to the SiO$_2$ surface. The liquid in the meniscus is represented by a meniscus resistance, $R_\mathrm{men}$, in parallel with a geometric meniscus capacitance, $C_\mathrm{men}$. Up to a geometric factor, they scale as $R_\mathrm{men}\sim\rho/r_\mathrm{men}$ and $C_\mathrm{men}\sim\varepsilon_0\varepsilon_\mathrm{w}r_\mathrm{men}$,
where $\rho$ is the resistivity of the liquid and $\varepsilon_\mathrm{w}\approx 80$ as a first approximation. The geometric factor cancels in the product,  $R_\mathrm{men}C_\mathrm{men}=\rho\,\varepsilon_0\varepsilon_\mathrm{w}$,
so at 100~kHz the meniscus behaves resistively for $\sigma\gtrsim\omega\varepsilon_0\varepsilon_\mathrm{w}\approx 4\times10^{-4}$~S~m$^{-1}$ and capacitively below. Even without added electrolyte, water equilibrated with air has $\sigma\approx 2\times10^{-4}$~S~m$^{-1}$,
which is close to this crossover value. Under nanoscale confinement near charged surfaces, the conductivity may be expected to be increased, to about $10^{-3}$~S m$^{-1}$ for $\sim$30~nm and about 0.1~S~m$^{-1}$ for $\sim$3~nm \cite{Wang2025}. Together with ions generated electrochemically during operation, this would suggest that the meniscus behaves resistively at 100~kHz.

The tip-meniscus interface is represented by an electric double layer (EDL) capacitance, $C_\mathrm{EDL,tip}$, in parallel with a charge-transfer resistance, $R_\mathrm{ct,tip}$. Similarly, the meniscus-CNT interface is represented by $C_\mathrm{EDL,CNT}$ in parallel with $R_\mathrm{ct,CNT}$, and the wetted SiO$_2$ surface by $C_\mathrm{EDL,SiO_2}$. All EDL capacitances are estimated as $C_{\mathrm{EDL},i}\approx\varepsilon_0\varepsilon_\mathrm{r}'A_i/h_\mathrm{EDL}$, with $i\in\{\mathrm{tip},\mathrm{CNT},\mathrm{SiO_2}\}$, $\varepsilon_\mathrm{r}'\approx 8$ and $h_\mathrm{EDL}\approx 1$~nm,
corresponding to a compact layer of reduced permittivity with about $7~\mu\mathrm{F\,cm^{-2}}$. We estimate the wetted areas as $A_\mathrm{tip}\approx 2\pi r_\mathrm{tip}^2$, $A_\mathrm{SiO_2}\approx\pi r_\mathrm{men}^2$, and $A_\mathrm{CNT}\approx 2\pi r_\mathrm{CNT}\cdot 2r_\mathrm{men}$ (an upper bound, since the underside of the CNT is in contact with the oxide). Using $r_\mathrm{men}=r_\mathrm{tip}$ and $r_\mathrm{CNT}=1$~nm, this gives $C_\mathrm{EDL,tip}\approx 0.3$~fF, $C_\mathrm{EDL,SiO_2}\approx 0.1$~fF and $C_\mathrm{EDL,CNT}\approx 20$~aF. From the meniscus, there are two parallel routes to the Si. (1)~The direct route crosses the oxide under the meniscus. It is represented by $C_\mathrm{SiO_2}=\varepsilon_0\varepsilon_\mathrm{ox}\pi r_\mathrm{men}^2/h_\mathrm{SiO_2}\approx 0.2$~aF in series with $C_\mathrm{EDL,SiO_2}$. Because $r_\mathrm{men}\ll h_\mathrm{SiO_2}$, fringing fields probably raise $C_\mathrm{SiO_2}$ to a few aF. (2)~The second route goes through the CNT. The CNT has no ohmic contact to either electrode: it is a floating electrode, coupled to the liquid through $C_\mathrm{EDL,CNT}$ and to the Si along its entire length $L_\mathrm{CNT}$ through $C_\mathrm{CNT-Si}=2\pi\varepsilon_0\varepsilon_\mathrm{eff}L_\mathrm{CNT}/\ln(2h_\mathrm{SiO_2}/r_\mathrm{CNT})\approx 20$~aF/µm. Here $\varepsilon_\mathrm{eff}\approx(\varepsilon_\mathrm{ox}+1)/2\approx 2.5$ accounts for the CNT lying at the SiO$_2$/air interface.
For $L_\mathrm{CNT}$ of tens of µm, this gives $C_\mathrm{CNT-Si}\approx 0.2-1$~fF, about $10^2-10^3$ times $C_\mathrm{SiO_2}$. Finally, the tip and cantilever couple directly to the Si through air, via $C_\mathrm{tip-Si}$.
We estimate this at a few aF for the apex alone, and possibly fF or more once the rest of the tip and the cantilever are included. At 100~kHz, the impedance of each capacitive element is $|Z|=1/(\omega C)$ with $\omega=2\pi\times 100$~kHz. Two elements are therefore negligible. First, $C_\mathrm{Si-platen}$ ($|Z|\approx 270~\Omega$) is effectively a short circuit, so the Si and platen form a single node. Second, $C_\mathrm{EDL,SiO_2}$ is about $10^2$ times larger than $C_\mathrm{SiO_2}$, with which it is in series. We further expect the parallel $C_\mathrm{men}$ and $R_\mathrm{men}$ to provide negligible impedance compared to the various liquid-solid interfaces.

\section{Micro-Raman spectroscopy and spectral analysis}

All Raman spectra were acquired confocally in backscattering geometry at $\lambda_\mathrm{laser}=532~\mathrm{nm}$ with linear polarization parallel to the CNT axis unless otherwise noted. Measurements in Ar were performed with the sample housed in an environmental stage (Linkam), using a $50\times$ objective (0.50 NA) and typically 1~mW on the sample. Water immersion measurements were performed with a $60\times$ water-immersion objective (1.00 NA). Line scans were acquired by translating the sample beneath the objective with a motorized stage (M\"arzh\"auser), stepping the laser spot along the CNT axis in increments of 1 or 2~µm.

Where the measured Si peak deviated from $520.7~\mathrm{cm}^{-1}$, spectra were first rigidly shifted to place it at this value. The RBM frequency $\omega_\mathrm{RBM}$ and full width at half maximum $w$ were then extracted by fitting a Lorentzian profile. Fit reliability was assessed using three criteria. First, the normalized root-mean-square error was calculated within 
$\omega_0 \pm w$, where $\omega_0$ is the fitted peak center and $w$ is the the fitted full width at half maximum (FWHM):
\begin{equation}
\mathrm{NRMSE}=\frac{1}{A}\sqrt{\frac{1}{N}\sum_{i=1}^{N}\left(I_i^{\mathrm{raw}}-
I_i^{\mathrm{fit}}\right)^2}.    
\end{equation}
Here, $A$ is the fitted Lorentzian amplitude. The relative FWHM uncertainty was defined as
\begin{equation}
\delta_w=\frac{\sigma_w}{w},    
\end{equation}
and the signal-to-noise ratio as
\begin{equation}
\mathrm{SNR}=\frac{A}{\sigma_{\mathrm{noise}}},    
\end{equation}
where $\sigma_{\mathrm{noise}}$ is the standard deviation of the fitting residuals outside the region $|\omega-\omega_0|\le 2w$. Only fits satisfying $\mathrm{NRMSE}<1$, $\frac{\sigma_w}{w}<1$, and $\mathrm{SNR}>2$ were retained.

The marker size in the plots indicates the SNR of the fitted RBM peaks. Larger markers represent spectra with $\mathrm{SNR} > 5$, whereas smaller markers correspond to spectra with $\mathrm{SNR}\leq 5$. Marker colors indicate the FWHM, and vertical error bars represent uncertainties in the fitted peak positions.

\section{AFM data of CNTs 2-4}
\label{S:AFM}

\begin{figure}[H]
    \centering
    \includegraphics[width=6.447in]{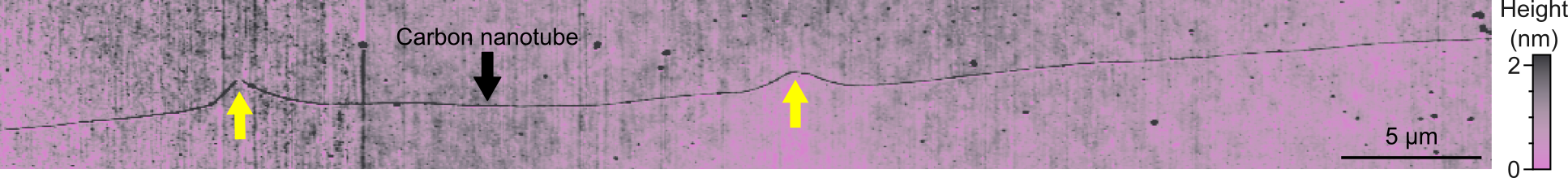}

    \caption{Tapping-mode AFM image of CNT~2 acquired at a relative humidity of $\mathrm{RH}=50\%$. Yellow arrows indicate LAO cut sites.}
    \label{fig:S2}
\end{figure}

\begin{figure}[H]
    \centering
    \includegraphics[width=6.436in]{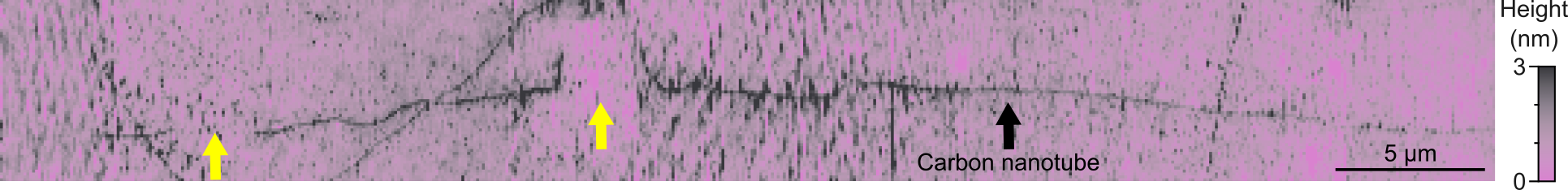}

    \caption{Tapping-mode AFM image of CNT~3 acquired at a relative humidity of $\mathrm{RH}=30\%$. Yellow arrows indicate LAO cut sites.}
    \label{fig:S3}
\end{figure}

\begin{figure}[H]
    \centering
    \includegraphics[width=6.387in]{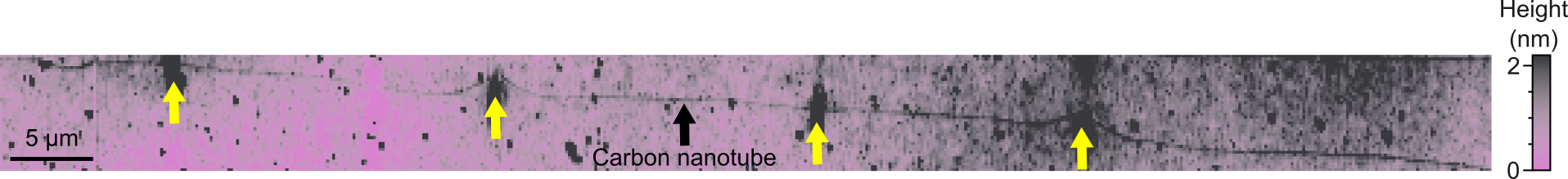}

    \caption{Tapping-mode AFM image of CNT~4 acquired at a relative humidity of $\mathrm{RH}=70\%$. Yellow arrows indicate LAO cut sites.}
    \label{fig:S4}
\end{figure}

Large-area tapping mode AFM images of CNTs 2-4 are shown in Figs.~\ref{fig:S2}-\ref{fig:S4}. The topography reveals the overall morphology and trajectory of each CNT across the substrate. Small kinks and lateral displacements are visible near some of the LAO-cut positions, which indicates slight mechanical dragging of the CNT during the cutting process. Typically, an unsuccessful first LAO cut attempt would result in such local displacement which may be rationalized given the relatively high force of $3$~µN used in this work, despite the feedback loop being on during LAO cutting. Two such LAO cut sites of CNT~2 are indicated in Fig.~\ref{fig:S2}. These correspond to the first two (leftmost) cuts indicated in Figs.~2(c) and \ref{fig:S6}b. In Figs.~\ref{fig:S3} and \ref{fig:S4} we similarly indicate LAO cut positions for CNTs 3 and 4, respectively. In the case of CNT~4, we systematically measure locally increased height at the LAO cut sites after cutting which would indicate material deposition that warrants further investigation.

\section{Quantitative assessment of LAO-induced defect creation}

\begin{figure}[H]
    \centering
    \includegraphics[width=3.611in]{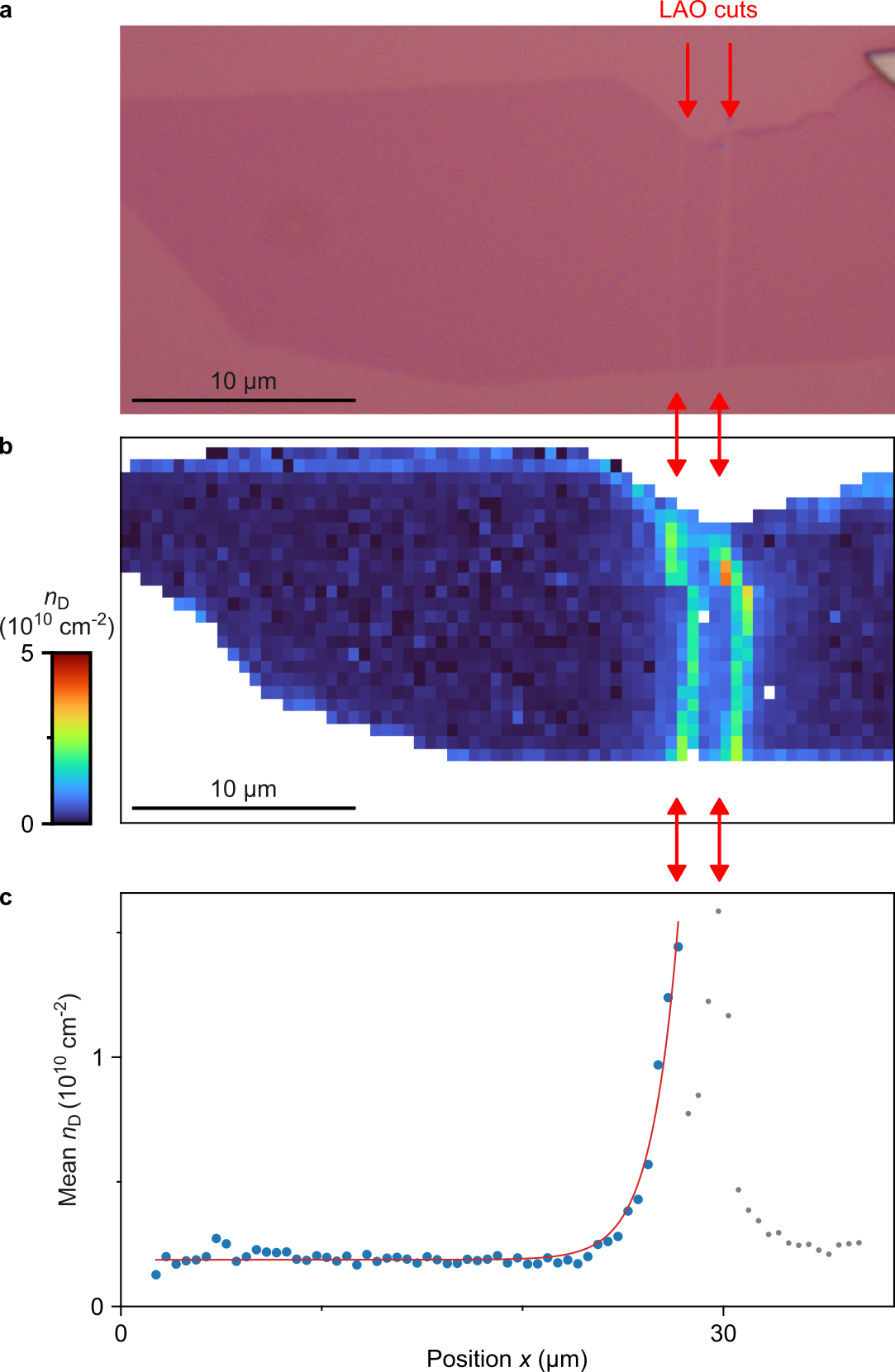}
    \caption{(a) Optical image of the monolayer graphene flake with two parallel LAO cuts indicated by the red arrows. (b) Corresponding spatial map of the defect density extracted from the $I(\mathrm{D})/I(\mathrm{G})$ ratio measured by micro-Raman spectroscopy at 532~nm laser excitation. (c) Average defect density along the direction perpendicular to the LAO cuts (blue data points) and fitted exponential trend line (red).}
    \label{fig:S5}
\end{figure}

To assess the spatial extent over which LAO cutting may induce defects in an sp$^2$-carbon lattice, we perform LAO cutting of a mechanically exfoliated monolayer graphene flake followed by quantitative micro-Raman spectroscopy mapping. The reason to do this on graphene rather than directly on a CNT is that models that quantitatively relate intensity ratios of Raman peaks, in particular of the D and G bands, to the areal density of defects are firmly established for graphene\cite{Cancado2011}, whereas for CNTs similarly general relations are not established \cite{Kalbacova2019,Sebastian2022}. Fig.~\ref{fig:S5}a shows an optical microscopy image of the flake where two parallel LAO cuts were performed across the flake at the locations indicated by the red arrows. LAO cutting was performed at a tip scan speed of
1~µm/s under a sinusoidal AC bias of $16~\mathrm{V_{pp}}$ at 100~kHz, with no DC offset and an applied normal force of
3~µN. The relative humidity was not recorded
during the experiment. Raman spectroscopy was performed after LAO cutting, and the intensities of the D and G bands were extracted. The $I_\mathrm{D}/I_\mathrm{G}$ ratio was then calculated and used to quantify the local defect density using the relation established by Can\c{c}ado \textit{et al.}~\cite{Cancado2011} for graphene in the low-defect-density regime:
\begin{equation}
L_\mathrm{D}^2~(\mathrm{nm}^2) = \left(1.8 \pm 0.5\right)\times 10^{-9}\lambda_\mathrm{L}^4 \left(\frac{I_\mathrm{D}}{I_\mathrm{G}}\right)^{-1}. 
\end{equation}
Here, $L_\mathrm{D}$ is the average distance between point defects and $\lambda_\mathrm{L}$ is the excitation wavelength in nanometers. The corresponding point-defect density is
\begin{equation}
n_\mathrm{D}~(\mathrm{cm}^{-2})=\frac{\left(1.8 \pm 0.5\right)\times 10^{22}}{\lambda_\mathrm{L}^4}\left(\frac{I_\mathrm{D}}{I_\mathrm{G}}\right).
\end{equation}
For the $532$~nm excitation used in our measurements, this relation becomes approximately
\begin{equation}
n_\mathrm{D}~(\mathrm{cm}^{-2})\approx2.25\times10^{11}\left(\frac{I_\mathrm{D}}{I_\mathrm{G}}\right).
\end{equation}

Fig.~\ref{fig:S5}(b) shows a spatial map of the defect density quantified in this way. The extracted defect density reaches approximately $2\times10^{10}\,\mathrm{cm^{-2}}$ near the cutting sites, whereas it remains on the order of $10^{9}\,\mathrm{cm^{-2}}$ in regions away from the cuts. Note that although the Raman D peak is activated also at the flake edges, the defect density is higher in vicinity of the LAO cuts compared to the flake edges. To quantify the spatial extent of increased defect density near the cuts, The defect density was averaged along the $y$ direction
within the selected graphene region, with the averaging
width varying with $x$ and reaching a maximum of
approximately $12\,\mu\mathrm{m}$.

The resulting profile was fitted over the range
$0$--$26\,\mu\mathrm{m}$ (Fig.~\ref{fig:S5}) using
\begin{equation}
\overline{n_\mathrm{D}}(d)=C+A\exp\left(-\frac{d}{L_{\mathrm{char}}}\right),    
\end{equation}
where $d$ is the distance from the reference endpoint in the decay direction, $A$ is the amplitude, $C$ is the background offset, and $L_{\mathrm{char}}$ is the characteristic decay length. The fit yields $L_{\mathrm{char}}=1.3\pm0.1$~µm, with the uncertainty representing the standard error of the fit.

The micrometer-scale characteristic decay length provides quantitative evidence that the elevated defect density is spatially localized near the LAO-cut regions rather than distributed across the graphene sheet. The low background density away from the cuts is consistent with the weak D-band response expected for high-quality mechanically exfoliated graphene. These observations support a localized reaction near the tip–sample meniscus. Although graphene and CNTs may respond differently to LAO, this control experiment suggests that widespread defect generation is unlikely to be the primary origin of the observed RBM blueshift and FWHM broadening observed in CNTs 2-5. In few-walled CNTs, like those selected in our study, it may further be expected that LAO causes lattice defects primarily in the outermost wall.

\section{Micro-Raman spectroscopy data of the G bands of CNTs~2-4}

\begin{figure}[H]
    \centering
    \includegraphics[width=5.705in]{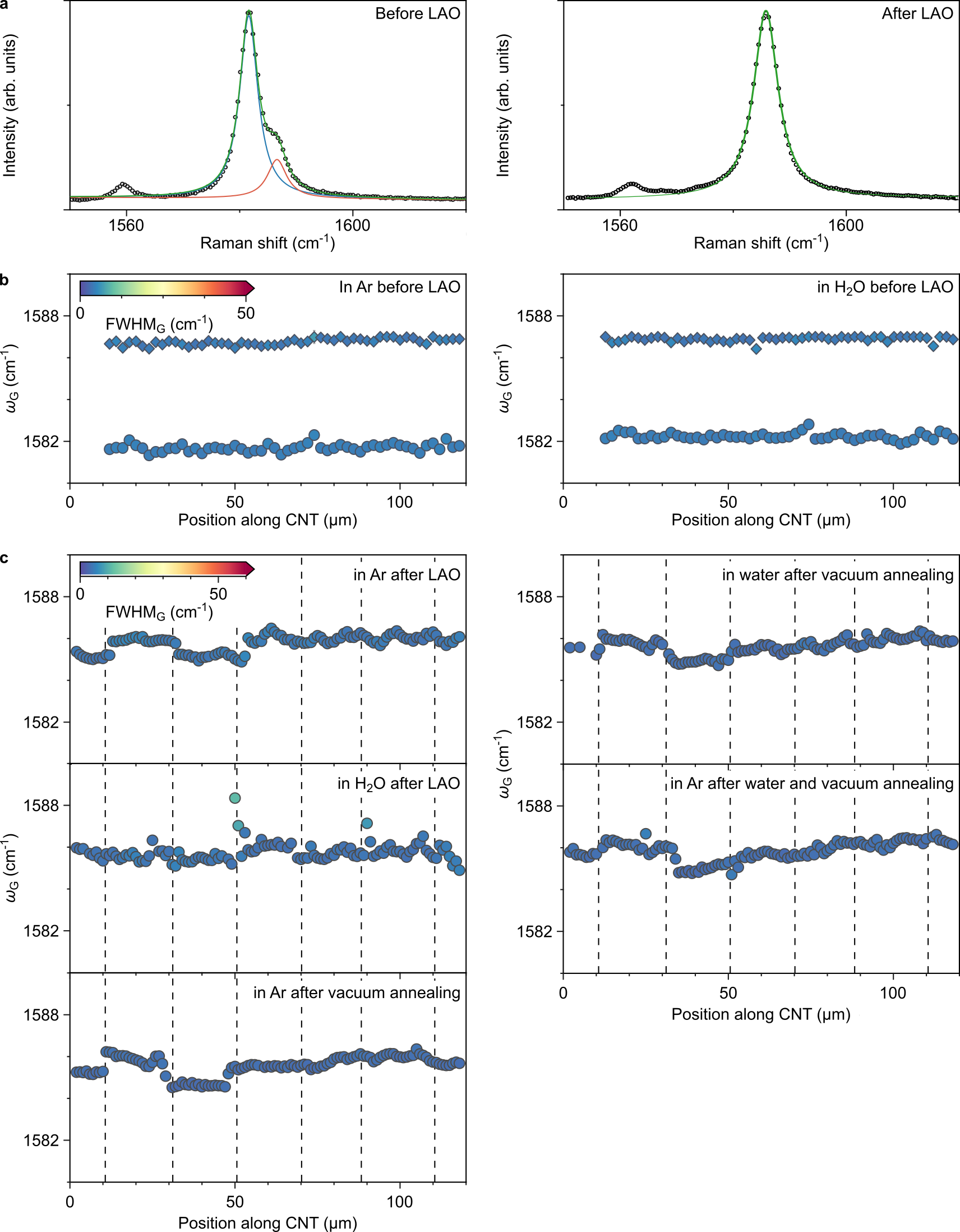}

    \caption{Micro-Raman spectroscopy of the Raman G band of CNT~2, measured along the same 120~µm long section studied in Fig.~2 of the main text. (a) Exemplary Raman G-band spectra before (left) and after (right) LAO. G-band frequency mapped vs position $x$ along the CNT (b) before and (c) after LAO.}
    \label{fig:S6}
\end{figure}

\begin{figure}[H]
    \centering
    \includegraphics[width=5.705in]{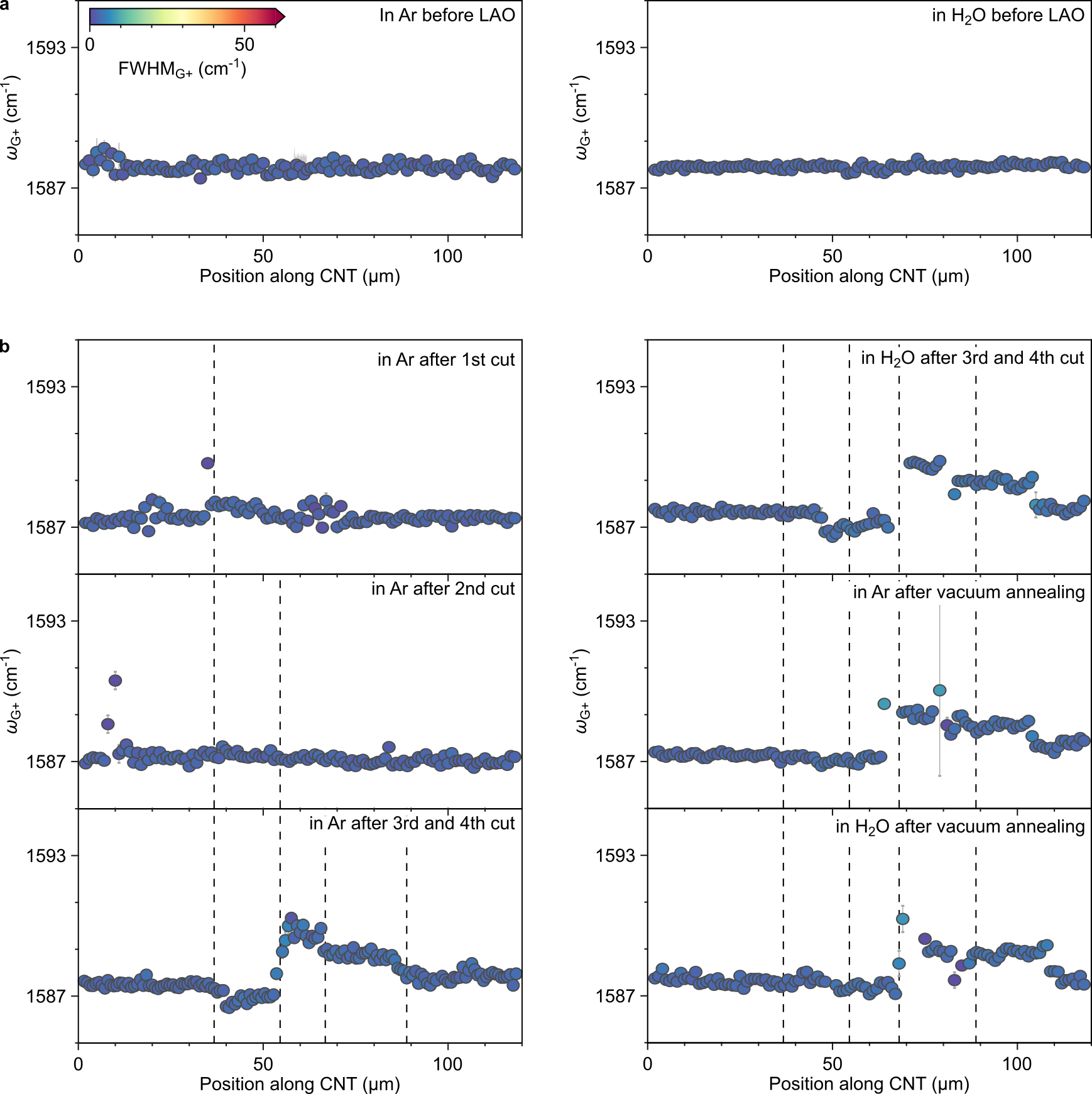}

    \caption{Micro-Raman spectroscopy of the Raman G band of CNT~3, measured along the same 120~µm long section studied in Fig.~3 of the main text (a) before and (b) after LAO.}
    \label{fig:S7}
\end{figure}

\begin{figure}[H]
    \centering
    \includegraphics[width=5.704in]{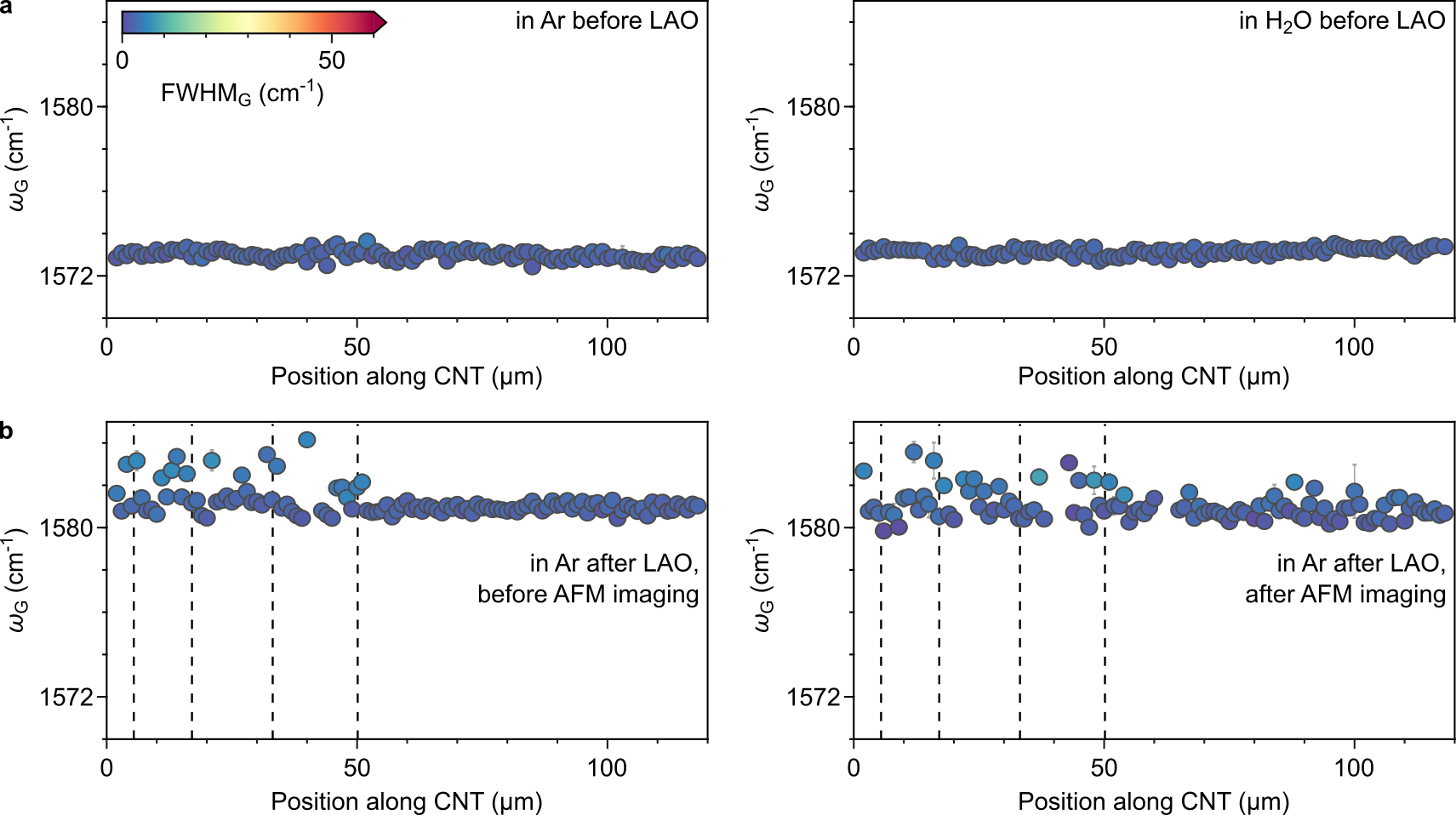}

    \caption{Micro-Raman spectroscopy of the Raman G band of CNT~4, measured along the same 120~µm long section studied in Fig.~4 of the main text (a) before and (b) after LAO.}
    \label{fig:S8}
\end{figure}

Figs.~\ref{fig:S6}-\ref{fig:S8} show Raman G-band data measured on CNTs~2–4 concurrently with the RBM data presented in Figs.~2-4 of the main text. We show G-band frequency, $\omega_\mathrm{G}$, and FWHM (color scale) as a function of position $x$ along each of the CNTs, typically of the most intense G-band component measured at 532~nm laser excitation unless otherwise noted. While we observe variations in G-band frequency as a function of sample history, as we discuss in the following, the G band does not show the same segment-specific response indicative of interior water filling that the RBM shows for each of these CNTs. Rather, it tends to show a distinct (from the RBM) spatial response characterized by what we interpret as (1) global changes in doping incurred between certain experiments \cite{Zhang2008,Pimonov2026}, and (2) more localized changes in axial strain due to LAO cutting-induced displacement \cite{Cronin2005}.

Fig.~\ref{fig:S6} shows micro-Raman spectroscopy data of the G band for CNT~2. For this CNT, we observe a change in G-band shape before and after LAO as illustrated in panel (a). We perform a double (single) Lorentzian fit of the higher intensity G-band component and show center frequencies vs position $x$ along the tube in panel b (c), respectively. Prior to LAO cutting, the G-band frequencies appear very uniform along the CNT, for both Ar and water environments. Following LAO cutting, a single high-intensity G-band component persists centered at $\sim1586~\mathrm{cm}^{-1}$, with fluctuations along the CNT length on the order of at most 2~cm$^{-1}$. There appear to be a few step-like changes at locations where LAO cuts have been performed, which we attribute to sub 1~\% variations in tensile strain \cite{Cronin2004,Cronin2005}.

Fig.~\ref{fig:S7} shows the frequency and FWHM of the strongest G-band component of CNT~3. As can be seen, before LAO cutting, $\omega_\mathrm{G}$ is very uniform along the entire 120~µm long section studied and centered at $\sim1588~\mathrm{cm}^{-1}$ both in Ar and under water immersion. The first two LAO cuts produce relatively small changes in the G-band frequency, whereas the third and fourth cuts induce a more pronounced blueshift over an extended region of the nanotube that is likely due to a change in tensile strain. The subsequent measurements in water and after vacuum annealing reveal additional changes in the spatial frequency distribution.

Fig.~\ref{fig:S8} shows the frequency and FWHM of the strongest G-band component of CNT~4. As can be seen, before LAO cutting, $\omega_\mathrm{G}$ is very uniform along the entire 120~µm long section studied and centered at $\sim1573~\mathrm{cm}^{-1}$ both in Ar and under water immersion. After LAO cutting, $\omega_\mathrm{G}$ still appears relatively uniform, but overall blueshifted to $\sim1581~\mathrm{cm}^{-1}$. This global shift might stem from a change in doping that could have possibly been introduced unintentionally during sample handling, or an unaccounted factor during LAO cutting that may equally have caused the increased height at LAO cut sites measured by AFM for this sample (Fig.~\ref{fig:S4}).

\section{Micro-Raman spectroscopy data of CNT~5}

\begin{figure}[H]
    \centering
    \includegraphics[width=5.705in]{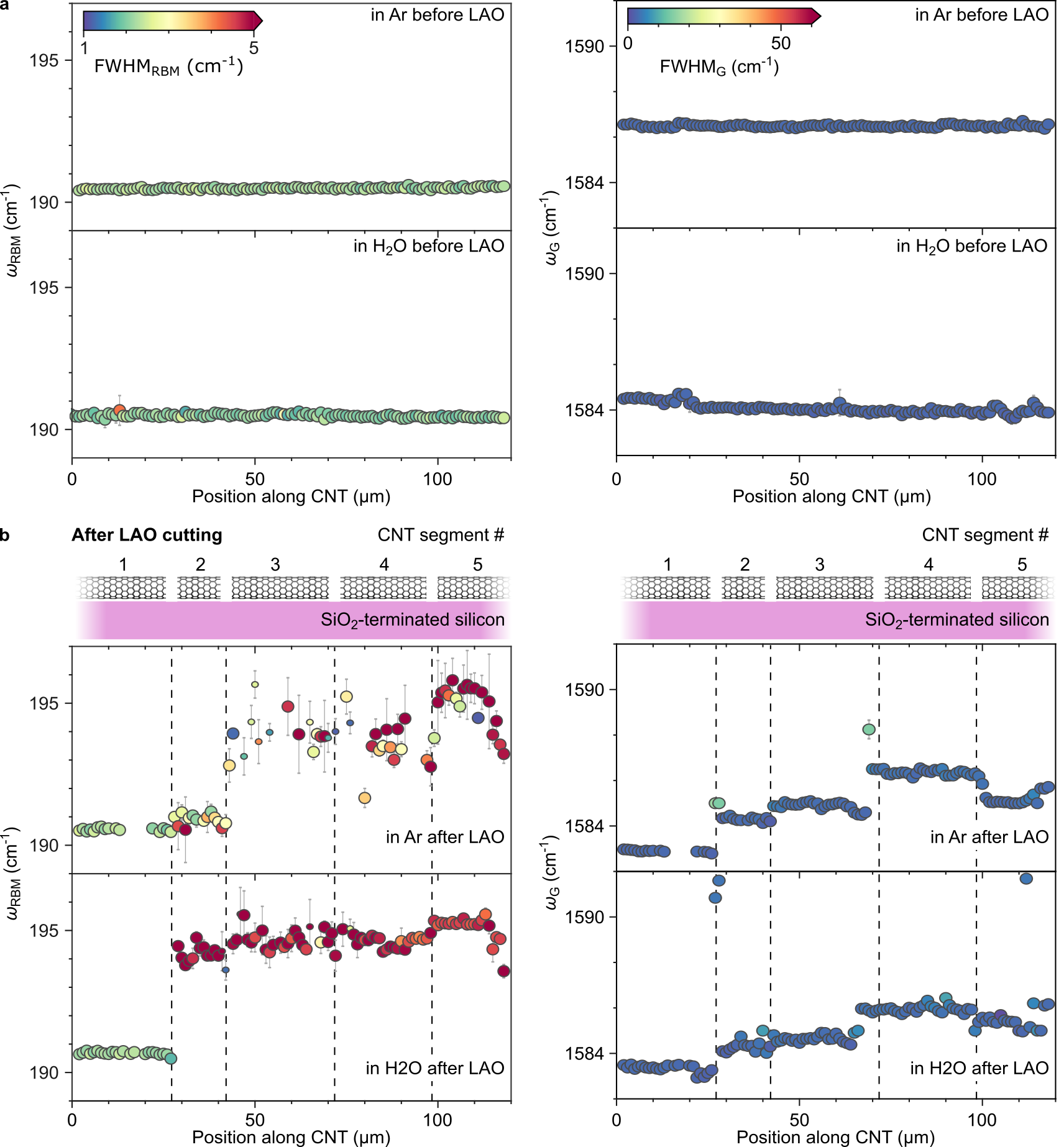}
    \caption{Micro-Raman spectroscopy of CNT~5 (a) before and (b) after LAO. Left (right) column shows the RBM (G) traces in different environments as indicated, and in chronological order from to top to bottom. Dashed black lines indicate approximate LAO cut positions.}
    \label{fig:S9}
\end{figure}

Fig.~\ref{fig:S9} shows micro-Raman spectroscopy data measured on an additional CNT (``CNT~5'') before and after LAO cutting. Similar to what we find for CNTs~2-4 in the main text, we observe that the highest frequency RBM, here at $\sim190.5~\mathrm{cm}^{-1}$ does not vary appreciably along the tube axis over the investigated 120~µm-long segment prior to LAO cutting. Further, $\omega_\mathrm{RBM}$ does initially not change upon water immersion of the CNT. After LAO cutting the tube at 4 locations along its axis (indicated by the dashed lines in Fig.~\ref{fig:S9}), we find RBM$_\mathrm{i}$ of segments 3, 4, and 5 to be upshifted by $\sim4.9~\mathrm{cm}^{-1}$, indicating interior filling. In contrast, RBM$_\mathrm{i}$ of segments 1 and 2 appear unshifted in Ar. Upon water immersion, we measure RBM$_\mathrm{i}$ of segment 2 to now shift to the same level as segments 3-5, indicating interior water filling of segment 2. In contrast, RBM$_\mathrm{i}$ of segment 1 remains unshifted at all times, indicating a closed segment.

\section{Meniscus geometry and electric field distribution}

Here we estimate whether the water meniscus condensed between the AFM tip and the substrate is plausibly taller than the CNT diameter, which favors LAO cutting because many AC cycles can then be completed while the meniscus connects the CNT and the tip, before the tip apex reaches the CNT (Fig.~\ref{fig:S1}). We further assess the resulting electric field distribution by finite element method (FEM) calculations.

\subsection{Zero-field meniscus}

We consider the simplest relevant geometry: a spherical Pt/Ir-coated AFM tip ($r_\mathrm{tip}=25~\mathrm{nm}$) in direct solid--solid contact with the flat SiO$_2$ substrate. At the point of contact, a condensed water meniscus takes the form of a small toroidal neck (a ``pendular ring'' in the classical capillary-bridge literature \cite{Orr1975}) wrapped around the contact point. Its curvature is set self-consistently by the Kelvin equation and the sphere--plane geometry. Using the surface tension of pure water $\gamma=72.8~\mathrm{mN/m}$ and its molar volume $V_\mathrm{m}=18\cdot10^{-6}~\mathrm{m^3/mol}$, the Kelvin equation yields an equilibrium (concave) curvature radius
\begin{equation}
r_\mathrm{K}=-\frac{\gamma V_\mathrm{m}}{RT\ln RH}\approx0.76\text{--}1.04~\mathrm{nm}
\label{eq:Kelvin}
\end{equation}
at $RH=50$--$60~\%$ and room temperature. Solving the exact sphere--plane meniscus geometry for this curvature assuming a circular arc meniscus profile yields a neck height of only $h\approx1.20$--$1.56~\mathrm{nm}$ and a liquid volume of $\approx108.59$--$183.41~\mathrm{nm^3}$ for dry surfaces. Because SiO$_2$ and the AFM tip are known to sustain a thin adsorbed wetting film below saturation, we repeat the calculation allowing the meniscus to merge smoothly onto such a film \cite{Crassous2011}. This enlarges the neck modestly, to $h\approx3.29$--$3.67~\mathrm{nm}$ and a volume of $\approx774.49$--$954.21~\mathrm{nm^3}$. In either case, the zero-field meniscus height remains small, at best comparable to, rather than significantly exceeding, typical CNT diameters.

\subsection{Field-enhanced capillary condensation}

An applied electric field is expected to stabilize a larger meniscus, as described by an electric field-modified version of the Kelvin equation \cite{Butt2011}. In the sign convention of Eq.~\ref{eq:Kelvin},
\begin{equation}
r_\mathrm{K,eff}=-\frac{\gamma}{\dfrac{RT\ln RH}{V_\mathrm{m}}+\dfrac{(\varepsilon_\mathrm{r}-1)\varepsilon_0E^2}{2}},
\label{eq:KelvinE}
\end{equation}
where $\varepsilon_0$ is the vacuum permittivity and $\varepsilon_\mathrm{r}$ is the relative permittivity of water. Eq.~\ref{eq:KelvinE} has been proposed for a flat capillary geometry, in which the electric field strength $E$ can be assumed homogeneous throughout the condensed liquid. In our case, however, $E$ is uniform neither spatially (throughout the meniscus) nor temporally (throughout one period of the sinusoidal AC bias). Deriving an electric field-modified Kelvin equation better suited to our geometry is beyond the scope of this work. It is further complicated by the strong and anisotropic modification of $\varepsilon_\mathrm{r}$ expected under nanoconfinement \cite{Fumagalli2018,Wang2025}, with anomalously low and even negative out-of-plane dielectric responses possible within the first 1--2 water layers near solid surfaces \cite{Bonthuis2011,Tran2023}, and by significant nonlinearities that arise at field strengths $E>1~\mathrm{V/nm}$ \cite{Butt2011}. For $E=0.5~\mathrm{V/nm}$ and the bulk value $\varepsilon_\mathrm{r}=80$, Eq.~\ref{eq:KelvinE} suggests $r_\mathrm{K,eff}\approx14.03~\mathrm{nm}$ at $RH=55~\%$, i.e., the electric field acts to stabilize a meniscus considerably larger than its zero-field size.

\subsection{Finite element calculations}

\begin{figure}
  \centering
  \includegraphics[width=2.907in]{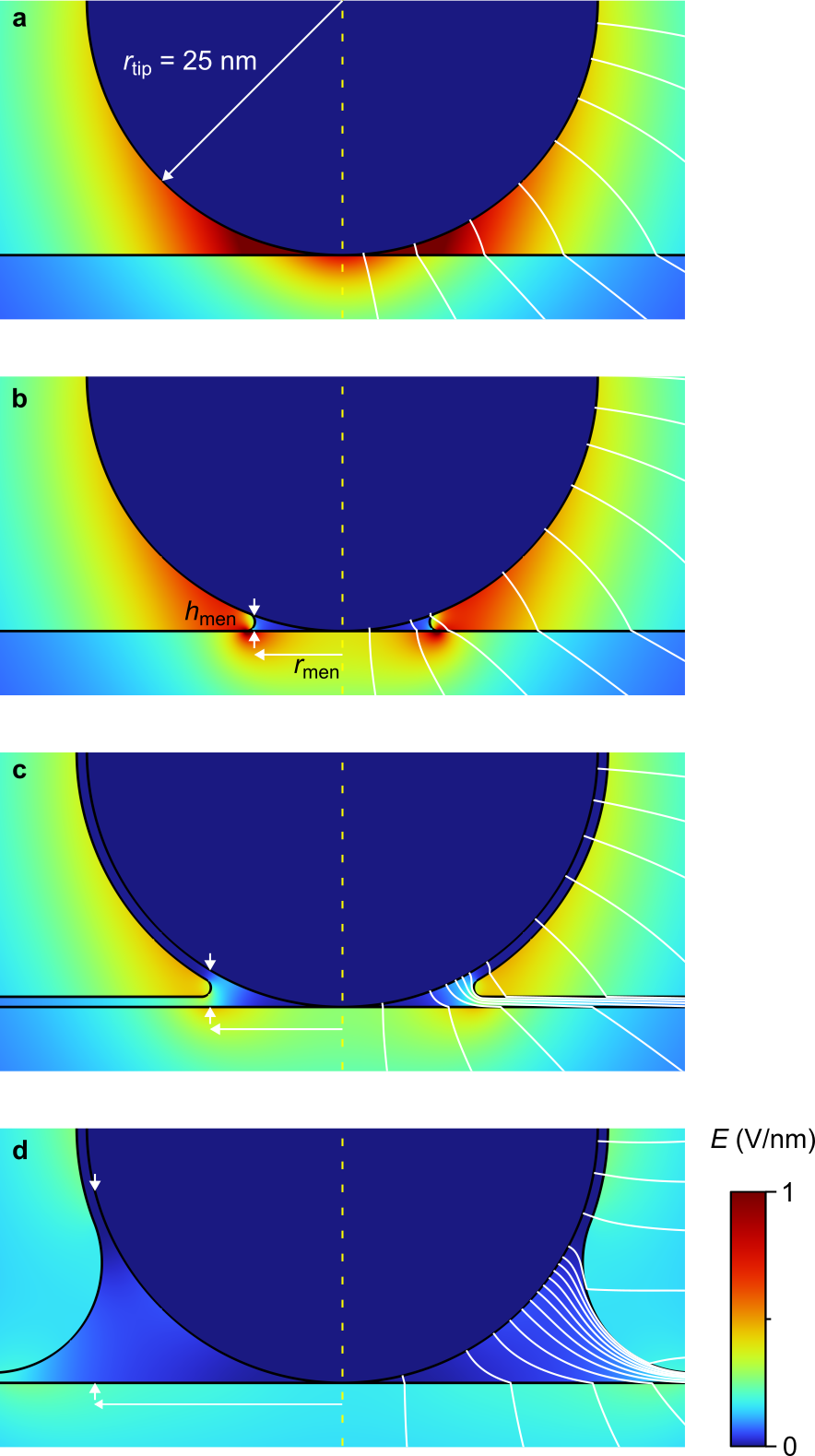}
  \caption{False-color plots of the electric field strength $E$ around the AFM tip obtained from finite element method (FEM) calculations at 8~V applied between the tip and the platen: (a) without water, (b) with capillary condensed water at $RH=55~\%$ and dry surfaces, (c) with capillary condensed water at $RH=55~\%$ and pre-wetted surfaces, and (d) with electric-field-enhanced capillary condensed water at $RH=55~\%$ and pre-wetted surfaces. Electric field lines are shown as white lines in the right half of each panel. For visual clarity, the color scale is truncated at an electric field strength of 1 V/nm.}
  \label{fig:S10}
\end{figure}

To gain further insight into the electric field distribution within the water meniscus, we performed FEM calculations (COMSOL), shown in Figure~\ref{fig:S10}. We consider a spherical AFM tip of radius $r_\mathrm{tip}=25~\mathrm{nm}$ in solid--solid contact with the SiO$_2$ surface at the tip apex, and an electrical potential difference of 8~V applied between this tip and an equipotential surface below the substrate. The substrate is modeled as a 300~nm thick SiO$_2$ layer with relative permittivity $\varepsilon_\mathrm{r}=3.75$.  The surrounding air ($\varepsilon_\mathrm{r}=1$) is modeled as an enclosing sphere much larger than the characteristic size of the system. For the water layers and the water meniscus, we use the bulk value $\varepsilon_\mathrm{r}=80$.

Although the FEM results are subject to the limitations discussed above, they allow a few general observations. In the absence of condensed water, the strongest electric field is localized near the tip apex (Figure~\ref{fig:S10}a). A condensed annular water meniscus instead defocuses the electric field and concentrates the strongest field in a ring around the tip apex, with a radius commensurate with the lateral extent of the meniscus (Figures~\ref{fig:S10}b--d). Such field focusing away from the tip apex agrees with previous FEM calculations and experiments on similar geometries \cite{Kremmer2003,Cambel2007}.

At $RH=55~\%$, representative of our experimental conditions, the meniscus height remains limited to $1.36~\mathrm{nm}$ as dictated by the Kelvin equation (Figure~\ref{fig:S10}b). It increases by only a factor of about 2--3 when pre-wetted surfaces are included, for which we use water layers of $1.0~\mathrm{nm}$ \cite{Crassous2011} (Figure~\ref{fig:S10}c). Under these conditions, the width of an LAO cut would be expected to be comparable to $r_\mathrm{tip}$, whereas AFM characterization after LAO cutting reveals larger cut widths on the order of $100~\mathrm{nm}$ (Section~\ref{S:AFM}). Absent a theory for electric-field-enhanced capillary condensation in our geometry, we estimate its impact on meniscus size and field distribution using $E=0.5~\mathrm{V/nm}$ in Eq.~\ref{eq:KelvinE} (Figure~\ref{fig:S10}d). Field enhancement plausibly stabilizes a meniscus large enough to rationalize the measured extent of our LAO cuts, but it also increases the defocusing of the electric field. Reaching field strengths on the order of 1~V/nm, as required to drive water splitting and carbon oxidation, would then clearly require additional ingredients. Anisotropic dielectric properties of water under confinement and near interfaces, as well as electric double layers at the solid--liquid interfaces, therefore appear key to a more complete description of the process \cite{Avouris1997,Bonthuis2011,Fumagalli2018,Tran2023,Wang2025}.

\section{EDX characterization of the AFM tip before and after LAO}

\begin{figure}[H]
    \centering
    \includegraphics[width=5.84in]{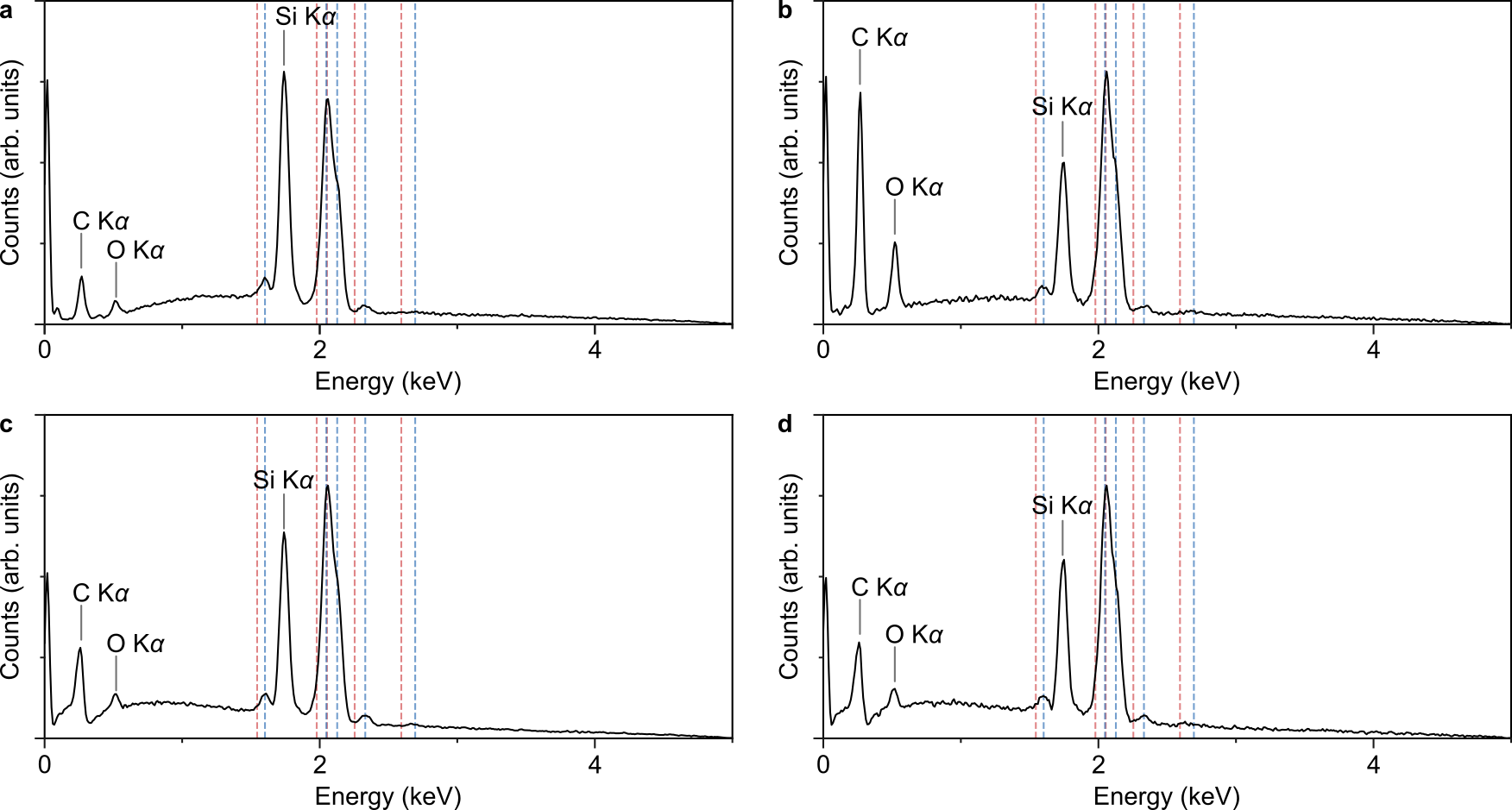}
    \caption{EDX results for the AFM tip for (a) the tip apex before LAO (b) the tip apex after LAO (c) the tip cantilever before LAO and (d) the tip cantilever after LAO. The red and blue dashed lines indicate the reference characteristic X-ray energies of iridium (Ir) and platinum (Pt), respectively.}
    \label{fig:S11}
\end{figure}

To investigate possible chemical changes induced by LAO, we performed energy-dispersive X-ray spectroscopy (EDX) on the AFM tip apex and cantilever before and after the cutting process. As shown in Figure~\ref{fig:S11}, the EDX spectra acquired at the tip apex following LAO exhibit an enhanced oxygen signal relative to the metal-related peaks. In contrast, the cantilever spectra show no comparably pronounced change in their relative elemental composition. These observations suggest that LAO induces localized oxygen enrichment near the tip apex, consistent with possible oxidation of the Ir/Pt coating during the cutting process.

\printbibliography